\documentclass[12pt,a4paper]{article}
\usepackage[T1]{fontenc}
\usepackage{amsmath,amssymb}
\usepackage{amsthm}
\usepackage[breaklinks]{hyperref}
\usepackage{verbatim}
\usepackage{color}
\usepackage{bm}
\usepackage{graphicx}
\usepackage{multirow}

\newtheorem{lemma}{Lemma}
\newtheorem{theorem}{Theorem}
\newtheorem{oldtheorem}{Theorem}

\newtheorem{corollary}{Corollary}
\newtheorem{claim}{Claim}

\numberwithin{subclaim}{claim}
\theoremstyle{definition}
\newtheorem{definition}{Definition}

\newcommand{\Accept}{\textsc{Accept}}
\newcommand{\atL}{\mathtt{l}}
\newcommand{\atR}{\mathtt{r}}

\newcommand{\set}[2]{\{ \, #1 \mid #2 \, \}}

\renewcommand{\emptyset}{\varnothing}
\renewcommand{\epsilon}{\varepsilon}
\renewcommand{\phi}{\varphi}

\newcommand{\rank}{\mathop{\mathrm{rank}}}

\newcommand{\haspath}{\mathop{\mathrm{HasPath}}}
\newcommand{\id}{\mathop{\mathrm{id}}}

\newcommand{\PL}[1]{\mathop{\mathrm{PL}_{#1}}}
\newcommand{\Player}[2]{\PL{#1}(#2)}
\newcommand{\Slayer}[2]{\mathop{\mathrm{SL}_{#1}(#2)}}

\newcommand{\lmark}{{\vdash}}
\newcommand{\rmark}{{\dashv}}

\DeclareRobustCommand{\stirling}{\genfrac\{\}{0pt}{}}

\begin{document}

\sloppy

\title{A lower bound on the state complexity of transforming two-way nondeterministic finite automata to unambiguous finite automata}
\author{Semyon Petrov\thanks{%
	Department of Mathematics and Computer Science,
	St.~Petersburg State University,
	14th Line V.~O., 29, Saint Petersburg 199178, Russia.
	E-mail: \texttt{semenuska2010@yandex.ru}
}
\and Alexander Okhotin\thanks{
	Department of Mathematics and Computer Science,
	St.~Petersburg State University,
	14th Line V.~O., 29, Saint Petersburg 199178, Russia.
	E-mail: \texttt{alexander.okhotin@spbu.ru}
}}

\maketitle

\begin{abstract}
This paper establishes a lower bound on the number of states 
necessary in the worst case to simulate
an $n$-state two-way nondeterministic finite automaton (2NFA)
by a one-way unambiguous finite automaton (UFA).
It is proved that for every $n$,
there is a language recognized by an $n$-state 2NFA
that requires a UFA with at least
$\sum_{k=1}^{n} (k - 1)! k! \stirling{n}{k} \stirling{n+1}{k}$ = $\Omega \big( n^{2n+2} / e^{2n} \big)$
states, where $\stirling{n}{k}$ denotes Stirling's numbers of the second kind.
This result is proved by estimating the rank of a certain matrix,
which is constructed for the universal language for $n$-state 2NFAs,
and describes every possible behaviour of these automata during their computation.

\end{abstract}

\section{Introduction}

A recurring question in finite automata theory
is the number of states in automata of one kind
needed to simulate an $n$-state finite automaton of another kind.
It remains unknown whether
nondeterministic two-way finite automata (2NFA) with $n$ states
can be simulated by deterministic two-way finite automata (2DFA) with polynomially many states,
and this question is known to be connected to the \emph{L vs.\ NL}
and \emph{L/poly vs.\ NL} problems in the complexity theory,
see Kapoutsis~\cite{Kapoutsis_logspace}
and Kapoutsis and Pighizzini~\cite{KapoutsisPighizzini_logspace}.
Much progress has been made on the complexity of simulating two-way automata by one-way automata.
The possibility of such a transformation
is one of the earliest results of automata theory~\cite{Shepherdson},
and the exact number of states sufficient and necessary to simulate 2NFA and 2DFA
by NFA and DFA
was later determined by Kapoutsis~\cite{Kapoutsis,Kapoutsis_thesis}.
Other related results include the complexity of transforming two-way automata to one-way automata
recognizing the complement of the original language,
first investigated by Vardi~\cite{Vardi}, followed by Birget~\cite{Birget_state_compressibility},
and later extended to alternating automata by Geffert and Okhotin~\cite{GeffertOkhotin_2afa}.
Attention was given to transforming two-way automata to one-way
in the case of alphabets of limited size:
in the unary case, the complexity was studied by Chrobak~\cite{Chrobak},
Mereghetti and Pighizzini~\cite{MereghettiPighizzini2001},
Geffert et al.~\cite{GeffertMereghettiPighizzini2003,GeffertMereghettiPighizzini}
and Kunc and Okhotin~\cite{TwoWayDFAs},
and later Geffert and Okhotin~\cite{GeffertOkhotin,GeffertOkhotin_2dfa_to_1dfa}
improved the bounds for alphabets of subexponential size.
Recently, the complexity of transforming sweeping permutation automata to one-way permutation automata
was determined by Radionova and Okhotin~\cite{RadionovaOkhotin}.

Among the results of Kapoutsis~\cite{Kapoutsis,Kapoutsis_thesis},
let us note the precise complexity of transforming 2NFA to one-way DFA and NFA.
For an $n$-state 2NFA,
the worst-case size of an equivalent DFA is $2^{n^2-O(n)}$ states~\cite{Kapoutsis_thesis},
whereas an equivalent NFA requires $\binom{2n}{n+1}$ states
in the worst case~\cite{Kapoutsis}.
Between these two perfectly conclusive results,
there is an open question involving an intermediate model between DFA and NFA:
the \emph{unambiguous finite automata} (UFA),
which can use nondeterminism, yet are bound to accept each string in at most one computation
(as in the unambiguous complexity classes, such as UL and UP).

State complexity of UFA is an active research topic.
Leung~\cite{Leung1998,Leung} studied relative succinctness of NFA with different degrees of ambiguity,
and, in particular, proved
that transforming an $n$-state UFA to a DFA requires $2^n$ states in the worst case~\cite{Leung}.
For the case of a unary alphabet, Okhotin~\cite{ufa_sc} showed
that the state complexity of transforming an $n$-state UFA to DFA
is of the order $e^{\Theta((\ln n)^{2/3})}$,
which was recently sharpened to $e^{(1+o(1))4^{1/3}(\ln n)^{2/3}}$ by F. Petrov~\cite{FPetrov_ufa}.
A notable fact is that the complement of every $n$-state UFA
can be recognized by a UFA with much fewer than $2^n$ states:
this was first proved by Jir\'asek Jr.\ et al.~\cite{JirasekjrJiraskovaSebej},
who demonstrated that at most $2^{0.79n}$ states are sufficient,
and soon thereafter Indzhev and Kiefer~\cite{IndzhevKiefer}
improved the upper bound to $\sqrt{n+1} \cdot 2^{n/2}$.
Lower bounds on the number of states necessary to represent the complement
have been researched as well:
in the case of a unary alphabet,
after an early lower bound $n^{2-\epsilon}$ given by Okhotin~\cite{ufa_sc},
Raskin~\cite{Raskin} discovered the state-of-the-art lower bound of $n^{(\log \log \log n)^{\Omega(1)}}$ states.
Later, G\"o\"os, Kiefer and Yuan~\cite{GoosKieferYuan}
used new methods based on communication complexity
to establish a higher lower bound of $n^{\Omega(\log n)/(\log\log n)^{O(1)}}$ states
for a two-symbol alphabet.
Most recently, Czerwi\'nski et al.~\cite{CzerwinskiDebskiGogaszHoiJainSkrzypczakStephanTan}
carried out a detailed analysis of descriptional and computational complexity of unary UFA,
and, in particular,
showed that every $n$-state unary UFA can be complemented using at most $n^{\log n+O(1)}$ states.

The complexity of transforming 2DFA to UFA was investigated recently by Petrov and Okhotin~\cite{LATA}.
Since UFA is an intermediate model between NFA and DFA,
this complexity has to lie between the two bounds of Kapoutsis~\cite{Kapoutsis}:
$\binom{2n}{n+1}$, which is of the order $\Theta\big(\frac{4^n}{\sqrt{n}}\big)$,
and $n(n^n - (n-1)^n)$.
Both bounds were improved.
An improved upper bound of $2^n \cdot n!$ states
was obtained by a new 2DFA-to-UFA transformation~\cite{LATA},
which augments the NFA constructed by Kapoutsis~\cite{Kapoutsis}
to ensure the uniqueness of its accepting computation.
A lower bound of $\Omega(5.656^n)$ states
was proved by reducing the task to estimating the rank of a certain matrix~\cite{LATA}.
The key element in the estimation is the rank of a particular submatrix,
which was earlier determined by Raz and Spieker~\cite{RazSpieker}.
Using it in the calculation leads to 
an improved lower bound of the order $9^n \cdot n^{-3/2}$
on the complexity of the 2DFA-to-UFA transformation~\cite{SPetrovFPetrovOkhotin}.

The complexity of the 2NFA-to-UFA transformation has not been investigated up to date.
It has to lie between the lower bound
of the order $9^n \cdot n^{-3/2}$
on the 2DFA to UFA transformation,
and Kapoutsis' bound of the 2NFA to DFA transformation,
which is 
of the order $2^{n^2 - O(n)}$.
It is natural to ask what is the exact function in the case of the 2NFA to UFA transformation;
these known bounds are even farther apart than in the 2DFA-to-UFA case.
This gap is narrowed in this paper
by improving the lower bound to
$\sum_{k=1}^{n} (k - 1)! k! \stirling{n}{k} \stirling{n+1}{k}$,
where $\stirling{n}{k}$ denotes Stirling's numbers of the second kind.
The new lower bound is asymptotically estimated as
$\Omega \big(\frac{n^{2n+2}}{e^{2n}}\big)$:
this is greater than the upper bound $2^n \cdot n!$
for the transformation of 2DFA to UFA,
showing that these two functions are definitely distinct.
Additionally, the new lower bound
is much closer to the upper bound $2^{n^2 - O(n)}$ than previously known.

\section{Definitions}

The paper uses standard finite automata models,
two-way and one-way,
nondeterministic of the general form and unambiguously nondeterministic.

For a set $X$, the set $\mathcal{P}(X)$ is defined as $\mathcal{P}(X) = \{ Y \mid Y \subseteq X \}$.

\begin{definition}
\label{definition_nfa}
A \emph{nondeterministic finite automaton} (NFA)
is a quintuple
$\mathcal{B}=(\Sigma, Q, Q_0, \delta, F)$,
in which
\begin{itemize}
\item
	$\Sigma$ is a finite alphabet;
\item
	$Q$ is a finite set of states;
\item
	$Q_0 \subseteq Q$ is the set of initial states;
\item
	the transition function
	$\delta \colon Q \times \Sigma \to \mathcal{P}(Q)$
	defines possible next states
	after reading a given symbol in a given state;
\item
	$F \subseteq Q$
	is the set of accepting states.
\end{itemize}

On an input string $w=a_1 \ldots a_\ell$,
a \emph{computation} is a sequence of states
$p_0, p_1, \ldots, p_\ell$
satisfying $p_0 \in Q_0$
and $p_{i+1} \in \delta(p_i, a_{i+1})$ for all $i$.
It is \emph{accepting} if, furthermore, $p_\ell \in F$.
The set of strings, on which there is at least one accepting computation,
is denoted by $L(\mathcal{B})$.
\end{definition}

An NFA is said to be \emph{unambiguous} (UFA),
if there is at most one accepting computation on each string.

Two-way automata operate on strings delimited by a left and a right end-marker.
\begin{definition}
\label{definition_2nfa}
A \emph{two-way nondeterministic finite automaton} (2NFA)
is a quintuple
$\mathcal{C}=(\Sigma, Q, Q_0, \delta, F)$,
in which
\begin{itemize}
\item
	$\Sigma$ is a finite alphabet,
	which does not contain a left end-marker ${\vdash} \notin \Sigma$
	and a right end-marker ${\dashv} \notin \Sigma$;
\item
	$Q$ is a finite set of states;
\item
	$Q_0 \subseteq Q$ is the set of initial states;
\item
	the transition function
	$\delta \colon Q \times (\Sigma \cup \{{\vdash}, {\dashv}\}) \to \mathcal{P}(Q \times \{-1, +1\})$
	defines possible transitions
	after reading a given symbol in a given state;
\item
	$F \subseteq Q$
	is the set of accepting states,
	effective at the right end-marker $\dashv$,
	with $\delta(q, {\dashv})=\emptyset$ for all $q \in F$.
\end{itemize}

On an input string $w=a_1 \ldots a_\ell$,
a \emph{computation} is a maximal sequence (which might be infinite) of pairs
$(p_0, k_0), (p_1, k_1), \ldots$,
where in each $i$-th pair $p_i \in Q$ is a state,
$k_i$ is the position of the head, with $0 \leqslant k_i \leqslant \ell + 1$. 
In addition, the following conditions should be satisfied: 
$p_0 \in Q_0$,
$k_0 = 0$,
and $(p_{i+1}, k_{i+1} - k_i) \in \delta(p_i, a_{k_i})$ for all $i$.
Here, $a_0 = {\vdash}$ and $a_{\ell + 1} = {\dashv}$.

Such pairs $(p_i, k_i)$ with $p_i \in Q$ and $k_i \geqslant 0$ are called \emph{configurations} of the automaton $\mathcal{C}$.
The pairs $(p_0, k_0)$ with $p_0 \in Q_0$ and $k_0 = 0$ are called \emph{initial configurations},
and a computation can start from any initial configuration.

The computation is \emph{accepting} if it is finite and its last pair
is $(p_n, k_n)$, with $p_n \in F$ and $k_n = \ell + 1$.
The set of strings, on which there is at least one accepting computation,
is denoted by $L(\mathcal{C})$.
\end{definition}

\section{Lower bound} \label{lower_bound_section}

\subsection{The witness language and its 2NFA}

Let $n$ be a positive integer.
The goal is to construct a witness language $L_n$,
which is recognized by a 2NFA using $n$ states,
but any equivalent UFA would require a substantial number of states.

Consider computations of a 2NFA on inputs of the form $\lmark xy \rmark$,
where $\lmark x$ is a \emph{prefix} and $y \rmark$ is a \emph{suffix}.
Every accepting computation of the 2NFA on this input, if one exists,
starts with its head inside the prefix on the symbol $\lmark$,
and ends with its head inside the suffix on the symbol $\rmark$.
Then, such a computation must move its head from the prefix to the suffix at least once.
The sequence of moves that cross the boundary between the prefix and the suffix
is known as the crossing sequence.
If all moves in the crossing sequence are known and fixed,
then the existence of the entire accepting computation can be checked
by collecting some information
on the computations that fit inside the prefix $\lmark x$ and the suffix $y \rmark$,
done separately for the prefix and for the suffix, 
and then checking that the results can be matched together.

A deterministic one-way automaton (DFA) simulating a 2NFA on $\lmark xy \rmark$,
after reading the prefix $x$,
must remember a fairly exhaustive set of computations of the 2NFA on the prefix,
because it cannot know in advance which of these computations later turn out to be necessary:
as shown by Kapoutsis, it requires $2^{n^2-O(n)}$ states in the worst case.
In the same situation, a nondeterministic one-way automaton (NFA)
may remember less information
by using its nondeterminism to remember only what will later be needed;
for this simulation, Kapoutsis proved that as few as $\binom{2n}{n+1}$ states are always sufficient.
However, in the latter construction,
several sequences of guesses may lead to the acceptance,
which causes a problem to an unambiguous automaton (UFA) doing the same check.
Intuitively, it should not be able to use its nondeterminism to the full extent,
and will end up using substantially more states.

The proposed witness language $L_n$
consists of all strings of the form $uv$,
where $u$ encodes a possible behaviour of an arbitrary $n$-state 2NFA on some prefix,
$v$ encodes its behaviour on some suffix,
and these two behaviours concatenated together allow at least one accepting computation.
This is in some sense a universal language for all $n$-state 2NFAs,
and it is recognized by a fixed $n$-state 2NFA.
This requires defining the encodings first,
and the first step is to understand what information needs to be included.

Consider a computation of a 2NFA on the input consisting of a prefix $\lmark x$ and a suffix $y \rmark$. 
Every configuration in this computation points at a particular symbol,
and hence the computation, as a sequence of configurations,
can be split into two kinds of segments:
those in which the 2NFA walks over the prefix,
and those in which it walks over the suffix.
All transitions made in a segment on the prefix depend only on $x$,
and all transitions made in a segment on the suffix depend only on $y$.

Every computation contains one or more segments on the prefix,
and the encoding for the prefix should describe
all such segments in all possible computations.
The first segment in a computation
begins in one of the starting states on the symbol $\lmark$,
and ends with the first crossing to the suffix;
all one has to remember about such segments is the set of states
in which the 2NFA can leave the prefix.
All other prefix segments are contained between two crossings
of the boundary between the prefix and the suffix,
one from $y \rmark$ to $\lmark x$, and the other from $\lmark x$ back to $y \rmark$.
It is sufficient to remember in which pairs of states the automaton can potentially make these two crossings
(the beginning and the end of a segment).

Similarly, every segment of an accepting computation on the suffix
is either a middle part contained between two crossings of the boundary,
or the final part that ends with acceptance.
For every possible middle part, it is sufficient to remember the states in which it crosses the boundary.
The final part is a computation that starts at the first symbol of $y \rmark$ in some state
and accepts the string,
and all one has to remember is the set of states from which one can reach acceptance.

Denote $[n] = \{1, 2, \dots, n\}$.
Both for the prefix and for the suffix,
all information on the possible middle segments
can be encoded as a function $f$ from $[n]$ to $\mathcal{P}([n])$:
for every state in the beginning of a segment (right after crossing the boundary),
it contains all possible resulting states in which the automaton can cross to the other side.
The initial segments (for a prefix), or the final segments (for a suffix)
are encoded as an extra subset of $[n]$:
for a prefix, these are the states on the boundary
reachable from an initial configuration,
and for a suffix, this is a set of states that lead to acceptance.
Both sets are not empty, for otherwise no string could be accepted.

Let $S$ be the set of states on the boundary reachable from the initial configuration 
without leaving the prefix in the meantime.
Assuming that computations leaving the prefix exist at all (that is, $S$ is not empty),
let $i$ be a state in which any of those computations first visits the last symbol of the prefix;
then, necessarily, $f(i) \subseteq S$, since each of the states in $f(i)$ 
can then be reached from the state $i$ without leaving the prefix in the meantime.
Additionally, one can pretend that the states in $S$
are also reachable from every possible state,
that is, $S \subseteq f(j)$ for every state $j$; 
after all, an ability to return to these states
would not let the automaton reach any states otherwise unreachable, 
as they are reachable from the initial configuration anyway.
With this modification, there is no need to remember $S$ at all,
because it equals the intersection of all $f(j)$ for all $j$.
Hence, the encoding of computations on a prefix
can be simplified down to the following definition
(which is a slight modification of the construction given by Kapoutsis~\cite[{\S}4.1]{Kapoutsis_thesis}).

\begin{definition}
A \emph{prefix table} is a function $f \colon [n] \rightarrow \mathcal{P}([n]) \setminus \emptyset$,
such that there exists at least one index $i$ with $f(i) \subseteq f(j)$ for all $j$.
Every such $i$ is called a \emph{starting state} of $f$.
\end{definition}

Note that if $i$ and $j$ are both starting states of $f$,
then $f(i) \subseteq f(j)$ and $f(j) \subseteq f(i)$; hence, $f(i) = f(j)$.

Denote as $s(f)$ the minimal integer that is a starting state of $f$.
Then, $f(s(f))$ equals the above set $S$.

For a prefix table $f$, its \emph{size} $|f|$ is calculated as $|f| = \sum_{i=1}^n |f(i)|$.

Two representations of prefix tables are frequently used in the following.
First, a prefix table $f$ can be considered as a matrix $M(f)$,
with $M(f)_{i,j} = 1$ if $j \in f(i)$, and $0$ otherwise.
The other representation is by a bipartite directed graph $G(f)$
on vertices $\{ \atL, \atR \} \times [n]$,
with $((\atL, i), (\atR, j)) \in E(G(f))$ if and only if $j \in f(i)$.

\begin{figure}[t]
\center{\includegraphics[scale=1]{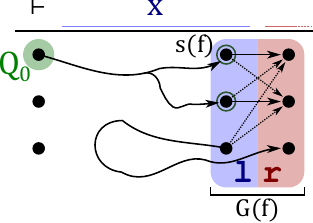}}
\caption{A prefix table constructed from an automaton on a prefix $x$}
\label{f:prefix_tables}
\end{figure}

The representation of computations by a prefix table is illustrated in Figure~\ref{f:prefix_tables}.
Possible transitions of the underlying 2NFA are depicted with solid arrows,
and the additional arcs are depicted as dotted arrows.
The resulting graph $G(f)$ is located on the right,
with the left part containing vertices of the form $(\atL, q)$,
and the right part containing vertices of the form $(\atR, q)$.
Starting states of $f$ are encircled.

The information about all computations on a suffix
is stored in a suffix table,
which, for every state $i$, represents all possible outcomes of the computations
beginning at the first symbol of a suffix in the state $i$:
the outcomes include crossing to the prefix in a certain state,
as well as acceptance.
If acceptance is among the possible outcomes,
then one can pretend that all other outcomes are possible as well:
indeed, if a string can already be accepted,
then no additional transitions could make it otherwise.
Naturally, acceptance must be a possible outcome for at least one state,
otherwise the string cannot be accepted at all.

\begin{definition}\label{definition_suffix_table}
A \emph{suffix table} is a function $g \colon [n] \rightarrow \mathcal{P}([n] \cup \{ \Accept \})$,
such that $\Accept \in g(i)$ for some $i$,
and for every index $i$, if $\Accept \in g(i)$, then $g(i) = [n] \cup \{ \Accept \}$.
\end{definition}

For a suffix table $g$, denote as $H(g)$ a bipartite directed graph
on the vertices $\{ \atL, \atR \} \times [n]$,
with $((\atR, i), (\atL, j)) \in E(H(g))$ if and only if $j \in g(i)$.
Also, define $A(g) = \{i \mid \Accept \in g(i) \}$;
this is the set of indices leading to acceptance.

\begin{figure}[t]
\center{\includegraphics[scale=1]{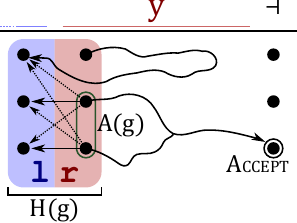}}
\caption{A suffix table constructed from an automaton on a suffix $y$.}
\label{f:suffix_tables}
\end{figure}

Computations on a suffix are represented by a suffix table
as illustrated in Figure~\ref{f:suffix_tables}.
Possible transitions of the underlying 2NFA are depicted with solid arrows,
and the arcs that were added are dotted.
The resulting graph $H(g)$ is located at the left of the figure,
with its left part containing vertices of the form $(\atL, q)$,
and the right part containing vertices of the form $(\atR, q)$.

Let $\Gamma_n = \{ (\atL, i) \mid i \in [n] \} \cup
\{ (\atL, f) \mid f \text{ is a prefix table} \} \cup
\{ (\atR, g) \mid g \text{ is a suffix table} \}$
be the alphabet of $L_n$, the universal language for $n$-state 2NFA.
This language is constructed as the language of a 2NFA $\mathcal{A}_n$ with the following properties:
the input alphabet is $\Gamma_n$;
the set of states is $[n]$;
the set of starting states is $\{ 1 \}$;
the set of accepting states is $[n]$.
It uses the transition function $\delta$, which is defined on $[n] \times \Gamma_n$ as follows:
\begin{itemize}
\item $\delta(q, (\atL, i)) = \{(i, +1)\}$;
\item $\delta(q, (\atL, f)) = f(q) \times \{ +1 \}$;
\item $\delta(q, (\atR, q)) = g(q) \times \{ -1 \}$ if $\Accept \notin g(q)$, and $\{(q, +1)\}$ otherwise.
\end{itemize}

If the input string is of a form $(\atL, s(f)) (\atL, f) (\atR, g)$,
for some prefix table $f$ and suffix table $g$,
then the automaton $\mathcal{A}_n$ first memorizes the starting state at the first symbol, and then follows the transitions in prefix and suffix tables, checking whether the accepting state is reachable. An alternate representation of the accepting computation is a path in the graph $G(f) \cup H(g)$ that starts in $(\atL, s(f))$ and ends in $(\atR, q)$ for some $q \in A(g)$; the vertices $(\atL, q)$ represent $\mathcal{A}_n$ being in state $q$ on the second symbol $(\atL, f)$, and the vertices $(\atR, q)$ represent $\mathcal{A}_n$ being in state $q$ on the third symbol $(\atR, g)$.

\begin{lemma} \label{graph_acceptance_lemma}
The automaton $\mathcal{A}_n$ accepts the string $(\atL, s(f)) (\atL, f) (\atR, g)$
if and only if
the graph $G(f) \cup H(g)$ has a path that starts in $(\atL, s(f))$ and ends in $(\atR, q)$ for some $q \in A(g)$.
\end{lemma}

\begin{proof}
If $\mathcal{A}_n$ accepts the string $(\atL, s(f)) (\atL, f) (\atR, g)$,
then a path can be constructed from an accepting computation as follows:
if $\mathcal{A}_n$ is on the symbol $(\atL, f)$ in state $q$, add the vertex $(\atL, q)$ to the path;
if $\mathcal{A}_n$ is on the symbol $(\atR, g)$ in state $q$, add the vertex $(\atR, q)$ to the path.
In both cases, the next transition corresponds to an arc from the current vertex in the graph
by definition of $G(f)$ ($(\atL, q)$ has arcs to all vertices $(\atR, i)$ with $i \in f(q)$)
and $H(g)$ ($(\atR, q)$ has arcs to all vertices $(\atL, i)$ with $i \in g(q)$).
The automaton $\mathcal{A}_n$ first visits the second symbol in the state $s(f)$ (dictated by the first symbol),
so the path starts in $(\atL, s(f))$;
since the computation is accepting, the last vertex on the path is $(\atR, q)$ with $\Accept \in g(q)$
(the automaton cannot get to an end-marker otherwise)---which is equivalent to $q \in A(g)$.

The converse statement has a similar proof: convert a path to a computation, and it will be accepting because at the end of the path it can leave the third symbol by moving to the right.
\end{proof}

The rest of this paper is concerned with proving a lower bound on the size of every UFA that recognizes the language $L_n$. To the best of our knowledge, the only known method for proving such lower bounds is the following theorem.

\begin{oldtheorem}[Schmidt~\cite{Schmidt}, see also Leung~\cite{Leung}]\label{Schmidt_theorem}
Let $L$ be a regular language,
and let $X$ and $Y$ be sets of strings.
Let $M$ be an integer matrix with rows labeled by elements of $X$ and columns labeled by elements of $Y$,
defined by $M_{x,y} = 1$, if $xy \in L$,
and $M_{x,y} = 0$ otherwise.
Then, every UFA for $L$ has at least $\rank M$ states.
\end{oldtheorem}

\begin{proof}[Sketch of a proof]
Let $X = \{x_1, x_2, \dots, x_l \}$, and let $Y = \{y_1, y_2, \dots, y_m\}$.

Let $U$ be a UFA that recognizes the language $L$.
Consider the matrix $M'$ 
with rows corresponding to the states of $U$
and columns corresponding to the strings $y_i$,
with values $M'_{q, y_i} = 1$ if $y_i \in L_q(U)$
($y_i$ is accepted from state $q$) 
and $M'_{q, y_i} = 0$ otherwise.

Note that every row of the matrix $M$ is a sum 
of some rows of the matrix $M'$; namely, of those that correspond to states
reachable from $q_0$ after reading $x_i$ (the sum in each cell cannot exceed 1, since $M$ is unambiguous).
Hence, $M'$ cannot contain fewer than $\rank M$ rows,
and the number of states of $U$ is at least $\rank M$.
\end{proof}

A string $(\atL, s(f)) (\atL, f) (\atR, g)$ can be naturally divided
into a prefix $(\atL, s(f)) (\atL, f)$ that depends only on a prefix table,
and a suffix $(\atR, g)$ that depends only on a suffix table.
The set of all such prefixes shall be used as $X$ in Schmidt's theorem,
and all such suffixes form the set $Y$.
Accordingly, the following matrix can be defined.

\begin{definition} 
The matrix $M^{(n)}$ is a matrix
with rows indexed by prefix tables and columns indexed by suffix tables,
with $M^{(n)}_{f,g} = 1$ if the automaton $\mathcal{A}_n$
accepts the string $(\atL, s(f)) (\atL, f) (\atR, g)$, and $0$ otherwise.
\end{definition}

In order to obtain a lower bound on the size of a UFA recognizing the language $L_n$ via Schmidt's theorem, 
the rank of this matrix needs to be calculated.

\subsection{Rank-preserving matrix reduction}

To calculate the rank of $M^{(n)}$, we shall simplify it first.
This can be done by identifying some prefixes of the form $(\atL, s(f)) (\atL, f)$,
for which the corresponding row of the matrix
is a linear combination of other rows.
Then, those prefix tables can be excluded
without affecting the rank of the resulting matrix.

For a bipartite directed graph $G(U, V, E)$ together with a vertex $s \in U$ and a set $A \subseteq V$, define the function $\haspath(G, s, A)$, which is equal to $1$ if there is a path in $G$ that starts in $s$ and ends in $A$, and $0$ otherwise. Note that, by Lemma~\ref{graph_acceptance_lemma}, the automaton $\mathcal{A}_n$ accepts the string $(\atL, s(f)) (\atL, f) (\atR, g)$ if and only if $\haspath(G(f) \cup H(g), (\atL, s(f)), \{ \atR \} \times A(g)) = 1$.

The redundancy of some rows in the matrix $M^{(n)}$ is ultimately inferred from the following lemma.

\begin{lemma} \label{lesser_inclusion_exclusion_formula}
Let $G(U, V, E)$ be a directed bipartite graph. Let $s \in U$ be a vertex, and let $A \subseteq V$ be a set of vertices. Let $u_1, u_2 \in U$ and $v_1, v_2 \in V$ be such vertices that $(u_1, v_1), (u_2, v_2) \in E$, but the edges $e = (u_1, v_2), e' = (u_2, v_1)$ are not in $E$.

Then, $\haspath(G, s, A) + \haspath(G + e + e', s, A) = \haspath(G + e, s, A)  + \haspath(G + e', s, A)$.
\end{lemma}

\begin{proof}
Obviously, if $\haspath(G, s, A) = 1$, then all four values of the function are equal to $1$, and the equality holds. Also, if $\haspath(G + e + e', s, A) = 0$, then all four values are equal to $0$, and the lemma is proved as well.

Suppose now, that $\haspath(G, s, A) = 0$ and $\haspath(G + e + e', s, A) = 1$, but $\haspath(G + e, s, A)  + \haspath(G + e', s, A) \neq 0 + 1 = 1$. Then, the latter sum is equal to $0$ or $2$; therefore, $\haspath(G + e, s, A)$ and $\haspath(G + e', s, A)$ are either both equal to $0$ or both equal to $1$.

Consider those two cases.

\begin{enumerate}
\item $\haspath(G + e, s, A)$ and $\haspath(G + e', s, A)$ are both equal to 0.

Since $\haspath(G + e + e', s, A) = 1$, there is a path from $s$ to $A$ in the graph $G + e + e'$.
Let $P$ be a simple path between them.
This path does not exist in graphs $G + e$ and $G + e'$
(otherwise, $\haspath$ would be equal to 1), therefore, both edges $e$ and $e'$ are in $P$.

If $e$ goes before $e'$ in $P$,
then the whole part of the path between them can be replaced with the edge $(u_1, v_1) \in E$.
Similarly, if $e'$ comes first, then this part can be replaced with the edge $(u_2, v_2) \in E$.
In both cases, $\haspath(G, s, A) = 1$, which contradicts the assumption that $\haspath(G, s, A) = 0$.

The case of $e$ going before $e'$ in $P$
is illustrated in Figure~\ref{f:sum_of_paths}(a),
in which the dotted lines represent the path $P$,
and the path from $s$ to $A$ in $G$ is obtained by replacing the longer route from $u_1$ to $v_1$
with a single arc.

\begin{figure}[t]
\center{\includegraphics[scale=1]{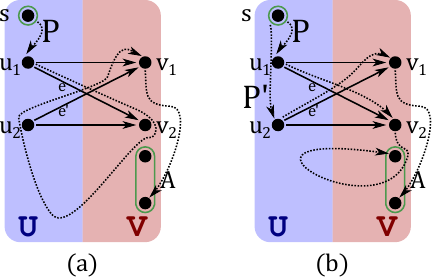}}
\caption{a) The combined path in the case $\haspath(G + e, s, A) = \haspath(G + e', s, A) = 0$.
b) Same, but in the case $\haspath(G + e, s, A) = \haspath(G + e', s, A) = 1$.}
\label{f:sum_of_paths}
\end{figure}

\item $\haspath(G + e, s, A)$ and $\haspath(G + e', s, A)$ are both equal to 1.

Let $P$ be a simple path from $s$ to $A$ in the graph $G + e$,
and let $P'$ be a simple path between them in $G + e'$. Note that $e \in P$ and $e' \in P'$, since $\haspath(G, s, A) = 0$. Then, the paths can be written as $P = P_1eP_2$ and $P' = P_1'e'P_2'$. However, the path $P_1 (u_1, v_1) P_2'$ contains neither $e$ nor $e'$, and is a path from $s$ to $A$; hence, $\haspath(G, s, A) = 1$, which leads to a contradiction.

Figure~\ref{f:sum_of_paths}(b) illustrates this case.
\end{enumerate}
\end{proof}

The same equality can be stated in terms of prefix tables. However, it requires proving that prefix tables augmented with one or two extra values remain valid prefix tables.

\begin{lemma} \label{lesser_inclusion_exclusion_formula_for_tables}
Let $f$ be a prefix table. Let $u_1 \neq u_2 \in [n]$ and $v_1 \neq v_2 \in [n]$ be such elements that $v_1 \in f(u_1), v_2 \in f(u_2)$, but $v_2 \notin f(u_1), v_1 \notin f(u_2)$.
Let $f_e$ be a function equal to $f$ except $f_e(u_1) = f(u_1) \cup \{ v_2 \}$; let $f_{e'}$ be a function equal to $f$ except $f_{e'}(u_2) = f(u_2) \cup \{ v_1 \}$; and let $f_{e+e'}$ be a function equal to $f$ except $f_{e+e'}(u_1) = f(u_1) \cup \{ v_2 \}$ and  $f_{e+e'}(u_2) = f(u_2) \cup \{ v_1 \}$.

Then, the functions $f_e$, $f_{e'}$ and $f_{e+e'}$ are prefix tables, and $M^{(n)}_{f,g} + M^{(n)}_{f_{e+e'},g} = M^{(n)}_{f_e,g} + M^{(n)}_{f_{e'},g}$ for every suffix table $g$.
\end{lemma}

\begin{proof}
First we shall prove that functions $f_e$, $f_{e'}$ and $f_{e+e'}$ are all prefix tables.
They all are defined on $[n]$, all their values are in $\mathcal{P}([n])$,
and all values are non-empty since they contain the corresponding value of $f$.
Then, it is enough to show that each of those functions has a starting state.
It is claimed that a starting state $s(f)$ of $f$ qualifies,
meaning that for all $j$, the conditions $f_e(s(f)) \subseteq f_e(j)$,
$f_{e'}(s(f)) \subseteq f_{e'}(j)$ and $f_{e+e'}(s(f)) \subseteq f_{e+e'}(j)$ hold true.

To prove this, it is enough to show that neither $u_1$ nor $u_2$ are the starting states of $f$;
indeed, if they are not, then $f_e(s(f)) = f(s(f)) \subseteq f(j) \subseteq f_e(j)$,
same for $f_{e'}$ and $f_{e+e'}$.
Suppose then, that $u_1$ is a starting state of $f$.
Then, $f(u_1) \subseteq f(u_2)$.
However, $v_1 \in f(u_1)$; then, $v_1 \in f(u_2)$ as well,
which leads to a contradiction since $v_1 \notin f(u_2)$.
The proof that $u_2$ is not a starting state is similar.
This confirms that $f_e$, $f_{e'}$ and $f_{e+e'}$ are prefix tables.

We would also want to prove that the minimal starting state
is the same for all four prefix tables;
that is, $s(f) = s(f_e) = s(f_{e'}) = s(f_{e+e'})$.
The proof is given for the case of $f_e$.
First, note that $f(s(f_e)) \subseteq f_e(s(f_e))$, because such a containment holds for every argument.
Next, $f_e(s(f_e)) \subseteq f_e(s(f))$, since $s(f_e)$ is a starting state of $f_e$.
The latter set $f_e(s(f))$ equals $f(s(f))$, because 
$s(f)$ is not equal to $u_1$ or to $u_2$,
and those are the only two indices for which the values of the four prefix tables differ.
And it is known that $f(s(f)) \subseteq f(j)$ for all $j$, because $s(f)$ is a starting state of $f$.
This chain of inclusions implies that $f(s(f_e)) \subseteq f(j)$ for all $j$,
meaning that $s(f_e)$ is a starting state of $f$.
Now, both $s(f)$ and $s(f_e)$ are starting states of both $f$ and $f_e$,
and hence, $s(f) \leqslant s(f_e)$ and $s(f_e) \leqslant s(f)$.
This proves that $s(f_e)=s(f)$.

It remains to show that $M^{(n)}_{f,g} + M^{(n)}_{f_{e+e'},g} = M^{(n)}_{f_e,g} + M^{(n)}_{f_{e'},g}$
for every suffix table $g$.
Let $G = G(f) \cup H(g)$.
Define $e = ((\atL, u_1), (\atR, v_2))$ and $e' = ((\atL, u_2), (\atR, v_1))$;
then, $G + e = G(f_e) \cup H(g)$, $G + e' = G(f_{e'}) \cup H(g)$ and $G + e + e' = G(f_{e+e'}) \cup H(g)$.

Then, by Lemma~\ref{lesser_inclusion_exclusion_formula},
$\haspath(G, (\atL, s(f)), \{ \atR \} \times A(g)) +
\haspath(G + e + e', (\atL, s(f)), \{ \atR \} \times A(g)) =
\haspath(G + e, (\atL, s(f)), \{ \atR \} \times A(g)) +
\haspath(G + e', (\atL, s(f)), \{ \atR \} \times A(g))$.
By Lemma~\ref{graph_acceptance_lemma}, each of those four values
can be represented in terms of the matrix $M^{(n)}$:
for instance, $\haspath(G, (\atL, s(f)), \{ \atR \} \times A(g)) =
\haspath(G(f) \cup H(g), (\atL, s(f)), \{ \atR \} \times A(g)) =
M^{(n)}_{f,g}$.
Since $s(f) = s(f_e) = s(f_{e'}) = s(f_{e+e'})$,
this transformation is valid for all four values.
The desired result then follows.
\end{proof}

Then, for any $f$ that satisfies the conditions
in Lemma~\ref{lesser_inclusion_exclusion_formula_for_tables},
the corresponding row $M^{(n)}_{f}$ in $M^{(n)}$
is a linear combination $M^{(n)}_{f_e} + M^{(n)}_{f_{e'}} - M^{(n)}_{f_{e+e'}}$
of rows for prefix tables of greater size,
and can therefore be removed without affecting the rank.
By iterating over all prefix tables $f$ in order of increasing size,
all remaining rows would correspond to prefix tables
for which the condition in Lemma~\ref{lesser_inclusion_exclusion_formula_for_tables}
does not hold.
The remaining prefix tables are called \emph{ordered}
(the meaning of this term will be revealed
in the next Section~\ref{lower_bound_section__ordered}).

\begin{definition}
A prefix table $f$ is called \emph{ordered} if there are no such elements $u_1 \neq u_2 \in [n]$ and $v_1 \neq v_2 \in [n]$ that $v_1 \in f(u_1), v_2 \in f(u_2)$, but $v_2 \notin f(u_1), v_1 \notin f(u_2)$.
\end{definition}

A reduced matrix can then be defined.

\begin{definition} 
The matrix $K^{(n)}$ is a matrix with rows indexed by ordered prefix tables and columns indexed by suffix tables, with $K^{(n)}_{f,g} =1$ if the automaton $\mathcal{A}_n$ accepts the string $(\atL, s(f)) (\atL, f) (\atR, g)$, and $0$ otherwise.
\end{definition}

In particular, $K^{(n)}_{f,g} = M^{(n)}_{f,g}$ for any ordered prefix table $f$ and any suffix table $g$.

\begin{figure}[t]
\center{\includegraphics[scale=2]{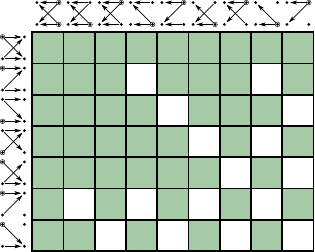}}
\caption{The matrix $K^{(2)}$, which is actually equal to $M^{(2)}$
(unordered prefix and suffix tables exist only for $n \geqslant 3$).}
\label{f:K_2}
\end{figure}

The matrix $K^{(n)}$ for $n = 2$ is illustrated in Figure~\ref{f:K_2}.
The rows are labeled by graphs corresponding to ordered prefix tables $f$,
with circled vertices corresponding to $s(f)$.
The columns are labeled by graphs corresponding to suffix tables $g$,
with circled vertices representing the ones that are in $A(g)$.
Filled cells are those containing 1, and empty cells contain zeroes.

\begin{theorem} \label{matrix_reduction_theorem}
$\rank K^{(n)} = \rank M^{(n)}$.
\end{theorem}

\begin{proof}
Construct a sequence of matrices $L_0 = M^{(n)}, L_1, \dots, L_m = K^{(n)}$ as follows: $L_k$ is a submatrix of $L_{k-1}$ obtained by removing a row corresponding to a non-ordered prefix table $f$ with minimal $|f|$. Since there is only a finite number of rows in $M^{(n)}$, this process will eventually stop; the resulting matrix will have rows corresponding to exactly all ordered prefix tables, so it is equal to $K^{(n)}$.

It is enough to prove that $\rank L_k = \rank L_{k-1}$ for all $k$; then all ranks of the matrices in the sequence are equal, and in particular, $\rank K^{(n)} = \rank L_m = \rank L_0 = \rank M^{(n)}$.

Let $f$ be the (non-ordered) prefix table corresponding to a row that is present in $L_{k-1}$ and absent in $L_k$. Since $f$ is not ordered, there exist indices $u_1 \neq u_2 \in [n]$ and $v_1 \neq v_2 \in [n]$ such that $v_1 \in f(u_1), v_2 \in f(u_2)$, but $v_2 \notin f(u_1), v_1 \notin f(u_2)$. Then, by Lemma~\ref{lesser_inclusion_exclusion_formula_for_tables}, $M^{(n)}_{f,g} + M^{(n)}_{f_{e+e'},g} = M^{(n)}_{f_e,g} + M^{(n)}_{f_{e'},g}$ for any suffix table $g$. Hence, $M^{(n)}_{f} = M^{(n)}_{f_e} + M^{(n)}_{f_{e'}} -  M^{(n)}_{f_{e+e'}}$ as rows. 

All of those rows are present in $L_{k-1}$, since $|f_e| = |f| + 1$, $|f_{e'}| = |f| + 1$ and $|f_{e+e'}| = |f| + 2$ are all greater than $|f|$, so the corresponding rows were not removed before the removal of $f$. Therefore, the row corresponding to a prefix table $f$ in the matrix $L_{k-1}$ can be represented as a linear combination of other rows in the same matrix. Hence, the rank of the matrix does not change upon deletion of such row, and  $\rank L_k = \rank L_{k-1}$.
\end{proof}

\subsection{Ordered prefix tables}\label{lower_bound_section__ordered}

There is a nicer characterization of a prefix table being ordered, which explain the choice of a term used.

\begin{lemma} \label{ordered_prefix_table_lemma}
A prefix table $f$ is ordered if and only if
for every two elements $i, j \in [n]$ either $f(i) \subseteq f(j)$ or $f(j) \subseteq f(i)$.
\end{lemma}

\begin{proof}
Assume $f$ is not ordered, but for every $i, j \in [n]$ either $f(i) \subseteq f(j)$ or $f(j) \subseteq f(i)$.
Then, there are elements $u_1 \neq u_2 \in [n]$ and $v_1 \neq v_2 \in [n]$
that $v_1 \in f(u_1), v_2 \in f(u_2)$, but $v_2 \notin f(u_1), v_1 \notin f(u_2)$.
Therefore, $f(u_1) \nsubseteq f(u_2)$, since $f(u_1)$ contains $v_1$, and $f(u_2)$ does not.
Similarly, $f(u_2) \nsubseteq f(u_1)$, since $f(u_2)$ contains $v_2$, but $f(u_1)$ does not.
However, either $f(u_1) \subseteq f(u_2)$ or $f(u_2) \subseteq f(u_1)$, which leads to a contradiction.

Conversely, assume that $f$ is ordered,
but there are elements $i, j \in [n]$ such that $f(i) \nsubseteq f(j)$ and $f(j) \nsubseteq f(i)$.
Obviously, $i \neq j$, otherwise $f(i) \subseteq f(j)$.
Let $u_1 = i$ and $u_2 = j$.
Let $v_1$ be any element of the set $f(i) \setminus f(j)$,
which is not empty since $f(i) \nsubseteq f(j)$.
Let $v_2 \in f(j) \setminus f(i)$; it is not empty because $f(j) \nsubseteq f(i)$.
Then, $v_1 \in f(u_1) = f(i), v_2 \in f(u_2) = f(j)$,
but $v_2 \notin f(u_1) = f(i), v_1 \notin f(u_2) = f(j)$,
which contradicts the table $f$'s being ordered.
\end{proof}

It follows that the values of an ordered prefix table $f$ can be ordered: $f(i_1) \subseteq f(i_2) \subseteq \dots \subseteq f(i_n) \subseteq [n]$. 

This chain of inclusions can contain identical sets, so an additional sequence is defined.
For an ordered prefix table $f$, let $S^f_0 \subsetneq S^f_1 \subsetneq S^f_2 \subsetneq \dots \subsetneq S^f_k = [n]$ be a sequence of unique sets in this ordering. Note that the set $[n]$ is always present, even if there is no index $i$ such that $f(i) = [n]$; this is relevant for the definition below.

\begin{definition}
The \emph{rank} of an ordered prefix table $f$, denoted $\rank f$, is equal to the number of unique sets $f(i) \neq [n]$.
\end{definition}

In the terms of a sequence $S^f_i$ above, the rank of $f$ is equal to $k$. 

The rank of $f$ is also equal to the rank of a matrix $J - M(f)$, where $J$ is a matrix in which every element is equal to $1$. In the matrix $J - M(f)$, the element on the intersection of row $i$ and column $j$ is equal to $1$ if $j \notin f(i)$, and $0$ otherwise.

\begin{lemma}
Let $f$ be an ordered prefix table. Then, $\rank f = \rank (J - M(f))$.
\end{lemma}
\begin{proof}
Let $\rank f = k$.
Consider the rows of a matrix $J - M(f)$. The intersection of a row $i$ and a column $j$ contains $1$ if $j \notin f(i)$, and $0$ otherwise; therefore, the entire row is uniquely determined by the value of $f(i)$. Moreover, if $f(i) = S_k^f = [n]$, the entire row contains only zeroes. Therefore, there are only $k$ possible non-zero values of rows in the matrix $J - M(f)$, corresponding to $f(i)$ being $S_0^f, S_1^f, \dots, S_{k-1}^f$ respectively. Hence, $\rank (J - M(f)) \leqslant k$.

The converse is proved by finding a full-rank $k \times k$ submatrix of $J - M(f)$.
Let $u_i$ for $0 \leqslant i \leqslant k - 1$ be some indices such that $f(u_i) = S_i^f$.
Let $v_j$ be some elements of $S_{j+1}^f \setminus S_j^f$ for $0 \leqslant j \leqslant k - 1$;
such indices always exist, since $S_0^f$ is non-empty and $S_{j-1}^f \subsetneq S_j^f$.

It is claimed that $(J - M(f))_{u_i, v_j} = 1$ if and only if $i \leqslant j$.
Indeed, $(J - M(f))_{u_i, v_j} = 1$ if and only if $v_j \notin f(u_i) = S_i^f$ by definition,
and the latter holds if and only if $i \leqslant j$,
because $v_j \notin S_j^{f}$ and $v_j \in S_{j+1}^f$ by the choice of $v_j$.
Thus, the submatrix of $J - M(f)$ formed by rows $u_0, \ldots, u_{k-1}$
and columns $v_0, \ldots, v_{k-1}$ is a upper-triangular matrix
with all 1$s$ on the main diagonal and above.
Hence, it is full-rank,
and the rank of the entire matrix $J - M(f)$ is at least $k$.

Since $\rank (J - M(f)) \leqslant k$ and $\rank (J - M(f)) \geqslant k$,
the values $\rank (J - M(f))$ and $k = \rank f$ are equal.
\end{proof}

\begin{definition}
For an ordered prefix table $f$ and an index $u \in [n]$,
the \emph{prefix layer} of $u$ in $f$, denoted by $\Player{f}{u}$,
is defined as the index $i$ such that $f(u) = S^f_i$.
\end{definition}

A prefix layer is a characterization of how powerful the given state is if obtained on a prefix.
All states on the same prefix layer lead to the same states on a suffix,
and a state on a higher prefix layer leads to a strictly larger set of states
than any state on a lower prefix layer.

\begin{figure}[t]
\center{\includegraphics[scale=1]{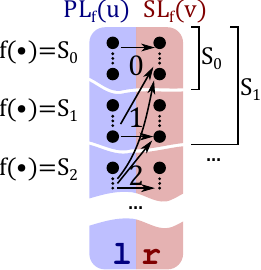}}
\caption{Prefix and suffix layers of a prefix table $f$.}
\label{f:layers}
\end{figure}

Note that the prefix layer of a starting state $s(f)$ is equal to 0,
and the prefix layer of every $u$ is bounded by the rank of the prefix table
($\Player{f}{u} \leqslant \rank f$).
All layers from $0$ up to $\rank f - 1$ must be non-empty,
whereas layer $\rank f$ may be empty
(it is empty if $[n]$ is not in the image of $f$).

\begin{definition}
For an ordered prefix table $f$ and an index $v \in [n]$,
the \emph{suffix layer} of $v$ in $f$, denoted by $\Slayer{f}{v}$,
is defined as the minimal index $i$ such that $v \in S^f_i$.
\end{definition}

In terms of the simulated automaton,
the suffix layer of a state indicates
how powerful a prefix state must be
to be able to get from it to this state on the suffix.

Prefix layers and suffix layers together form a structure of a given prefix table.
This structure is illustrated in Figure~\ref{f:layers}.

\subsection{Linear independence of rows of the matrix}

Rows of $K^{(n)}$ correspond to ordered prefix tables,
and are actually linearly independent.
The proof of this relies on finding a certain difference
between any two distinct ordered prefix tables.
This notion of difference is expressed in the definitions below.

\begin{definition}
Let $f_{\circ}$ be an ordered prefix table.
For another ordered prefix table $f$ and a layer $i$ ($0 \leqslant i < \rank f_{\circ})$,
the prefix table $f$ is said to \emph{break through} layer $i$ of $f_{\circ}$
if there are such indices $u, v \in [n]$ that $\Player{f_{\circ}}{u} \leqslant i$,
$\Slayer{f_{\circ}}{v} > i$ and $v \in f(u)$.
The set of all layers that $f$ breaks through for a given $f_{\circ}$ is denoted by $B_{f_{\circ}}(f)$.
\end{definition}

\begin{figure}[t]
\center{\includegraphics[scale=1]{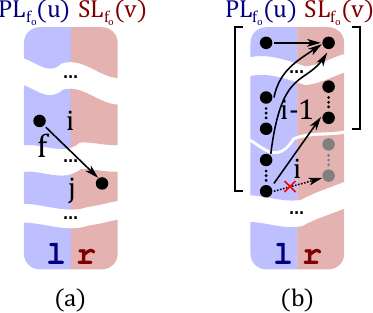}}
\caption{a) A prefix table that breaks through all layers between $i$ and $j - 1$ of $f_{\circ}$, inclusive.
b) A prefix table that drops down from layer $i$ of $f_{\circ}$.}
\label{f:break_through_layer}
\end{figure}

This notion is illustrated in Figure~\ref{f:break_through_layer}(a).

In other words, $f$ breaks through layer $i$
if it is stronger than $f_{\circ}$ with regard to this layer:
starting in all prefix states of all layers up to $i$ of $f_{\circ}$,
one can get to a suffix state on a higher layer by following the arrows,
regardless of the suffix table used.

The next lemma shows
that a breakthrough is guaranteed if $f$ has at least as many arcs as $f_{\circ}$
and is different from $f_{\circ}$.

\begin{lemma} \label{divergence_breakthrough_lemma}
If $f \neq f_{\circ}$ and $|f| \geqslant |f_{\circ}|$, then $B_{f_{\circ}}(f)$ is not empty.
\end{lemma}

\begin{proof}
Suppose that $B_{f_{\circ}}(f)$ is empty.
Then, for every $u \in [n]$ with $\Player{f_{\circ}}{u} = i$,
all elements $v \in f(u)$ have $\Slayer{f_{\circ}}{v} \leqslant i$,
and thus $f(u) \subseteq S^{f^{\circ}}_{i} = f_{\circ}(u)$.
In particular, $|f(u)| \leqslant |f_{\circ}(u)|$.
However, their sum $|f| = \sum_{u = 1}^n |f(u)|$ 
is at least $|f_{\circ}| = \sum_{u = 1}^n |f_{\circ}(u)|$ by assumption;
therefore, for every $u \in [n]$, the equality $|f(u)| = |f_{\circ}(u)|$ holds.
Since $f(u) \subseteq f_{\circ}(u)$ and $|f(u)| = |f_{\circ}(u)|$, it follows that $f(u) = f_{\circ}(u)$.
This is true for every $u \in [n]$, so $f = f_{\circ}$, which is a contradiction.
\end{proof}

\begin{definition}
Let $f_{\circ}$ be an ordered prefix table.
For an ordered prefix table $f$ and a layer $i$ ($0 \leqslant i < \rank f_{\circ})$,
the table $f$ is said to \emph{drop down} from layer $i$ of $f_{\circ}$
if for every $u \in [n]$ with $\Player{f_{\circ}}{u} \leqslant i$
and for all $v \in f(u)$, the suffix layer $\Slayer{f_{\circ}}{v}$ is strictly less than $i$.
\end{definition}

In other words, $f$ drops down from the layer $i$
if it is weaker than $f_{\circ}$ with regard to this layer:
starting in all prefix states of all layers up to $i$ of $f_{\circ}$,
and following the arrows provided by $f$,
one cannot get to a suffix state not only in layers beyond the $i$-th,
but even in the $i$-th layer itself.
This is illustrated in Figure~\ref{f:break_through_layer}(b).

Note that a prefix table does not break through or drop down from its own layers, due to how both definitions work.

\begin{theorem} \label{linear_independence_of_K_theorem}
Rows of $K^{(n)}$ are linearly independent.
\end{theorem}

\begin{proof}
Suppose this is not the case,
and there is a linear combination $\sum_{f} \alpha_f K^{(n)}_f = \mathbf{0}$,
where $\mathbf{0}$ is a vector of all zeroes,
the sum is over all ordered prefix tables,
and not all $\alpha_f$ are equal to $0$.
Among all ordered prefix tables $f$ with $\alpha_f \neq 0$,
let $f_{\circ}$ be one of those with minimal $|f_{\circ}|$,
that is, with the least number of arcs defined.

The general plan is to show that for each $f$ with $f \neq f_\circ$ and $\alpha_f \neq 0$,
the difference between $f$ and $f_\circ$
manifests itself in the linear combination,
so that there is a column that does not come down to 0 in the total sum.
By Lemma~\ref{divergence_breakthrough_lemma},
every such $f$ breaks through some layer of $f_\circ$
simply be the virtue of having at least as many arcs and being different;
whereas $f_\circ$ does not break through its own layers.

The first goal is to construct, for each ordered prefix table $f$,
a corresponding suffix table $g$,
such that,
in the graph formed by $f$ and $g$ together,
there is a path that passes through all the layers of $f_{\circ}$
and reaches acceptance,
whereas in the graph formed by $f_{\circ}$ and $g$ there is no such path.
This $g$ depends only on the set of breakthroughs $B_{f_{\circ}}(f)$ for $f$,
and what $g$ does is to stage additional breakthroughs in all remaining layers.

Let $k = \rank f_{\circ}$.
For each subset $I \subseteq \{0, 1, \dots, k-1\}$,
construct a suffix table $g_I$ as follows:
for every index $v$ with $\Slayer{f_{\circ}}{v} \notin I$,
the set $g_I(v)$ contains all nodes $u$ with $\Player{f_{\circ}}{u} \leqslant \Slayer{f_{\circ}}{v}$,
that is, no breakthroughs are attempted;
and for $v$ with $\Slayer{f_{\circ}}{v} \in I$,
the set $g_I(v)$ consists of all nodes $u$ with
$\Player{f_{\circ}}{u} \leqslant \Slayer{f_{\circ}}{v} + 1$,
thus ``breaking through'' the next layer.
The element
$\Accept$ is considered to be on the prefix layer $k$ for the purposes of this definition.
More formally,
\begin{equation*}
	g_I(v) = \begin{cases}
		\{ u \mid \Player{f_{\circ}}{u} \leqslant \Slayer{f_{\circ}}{v} \},
			& \text{if } \Slayer{f_{\circ}}{v} \notin I \text{ and } \Slayer{f_{\circ}}{v} < k
				\\
		\{ u \mid \Player{f_{\circ}}{u} \leqslant \Slayer{f_{\circ}}{v} + 1 \},
			& \text{if } \Slayer{f_{\circ}}{v} \in I \text{ and } \Slayer{f_{\circ}}{v} + 1 < k
				\\
		[n] \cup \{\Accept\},
			& \text{otherwise}
	\end{cases}
\end{equation*}

The desired properties of the suffix tables $g_I$
are established in the three claims below.
The first claim is that the set $A(g_I)$ of indices of $g_I$ leading to acceptance
is exactly the top $k$-th suffix layer of $f_\circ$,
and also its $(k-1)$-th suffix layer, provided that $g_I$ stages a breakthrough at layer $k-1$.
No elements from any lower layers may be in $A(g_I)$.

\begin{claim}\label{A_of_g_I_claim}
Let $f_{\circ}$ be an ordered prefix table, let $k$ be its rank,
and let suffix tables $g_I$ be constructed using $f_{\circ}$ as described above.
Then, if $k-1 \in I$, then $A(g_I) = \set{v}{\Slayer{f_{\circ}}{v} \geqslant k-1}$,
and if $k-1 \notin I$, then $A(g_I) = \set{v}{\Slayer{f_{\circ}}{v}=k}$.
\end{claim}
\begin{proof}
The set $A(g_I)$ is defined as a set of indices $v$
such that $g_I(v)$ contains $\Accept$,
which only happens in the definition above in the third case. 

If $\Slayer{f_{\circ}}{v} < k-1$,
then such $v$ will always satisfy one of the first two cases, so $v$ cannot be in $A(g_I)$.
If $\Slayer{f_{\circ}}{v} > k-1$,
then $g_I(v)$ is always defined as $[n] \cup \{\Accept\}$, and thus $v \in A(g_I)$.
Finally, if $\Slayer{f_{\circ}}{v} = k-1$,
then $g_I(v)$ will use the first case in the definition if $k-1 \notin I$ and the third case otherwise.

Thus, if $k-1 \notin I$, then $A(g_I)$ is equal to the set of indices $v$
for which $\Slayer{f_{\circ}}{v}$ is at least $k$.
Since $\Slayer{f_{\circ}}{v} \leqslant \rank f_{\circ} = k$,
this is equivalent to $\Slayer{f_{\circ}}{v}$ being exactly $k$,
and hence $A(g_I) = \set{v}{\Slayer{f_{\circ}}{v}=k}$, as claimed.
In the case that $k-1 \in I$,
the set $A(g_I)$ contains both indices $v$
with $\Slayer{f_{\circ}}{v} > k - 1$ and with $\Slayer{f_{\circ}}{v} = k - 1$, and nothing else;
this condition can be simplified as $\Slayer{f_{\circ}}{v} \geqslant k - 1$.
\end{proof}

The second claim is that if an ordered prefix table
drops down from any layer of $f_{\circ}$,
then this prefix table does not affect the linear combination
on the columns corresponding to $g_I$, for various $I$.

\begin{claim}\label{sustenance_failing_claim}
Let $f$ and $f_{\circ}$ be two ordered prefix tables, 
and let suffix tables $g_I$ be constructed using $f_{\circ}$ as described above.
If there is a layer $i$ of $f_{\circ}$ from which $f$ drops down,
then for every set $I$ the value of $K^{(n)}_{f,g_I}$ is equal to $0$.
\end{claim}
\begin{proof}
By Lemma~\ref{graph_acceptance_lemma},
$K^{(n)}_{f,g_I} = 0$ if and only if
there is no path in $G(f) \cup H(g_I)$
that starts in $(\atL, s(f))$
and ends in $(\atR, q)$ for some $q \in A(g_I)$.

It can be shown that any path from $(\atL, s(f))$
never goes beyond prefix layer $i$ on the left
and never reaches the $i$-th layer or beyond on the right;
formally,
it only visits vertices $(\atL, u)$ with $\Player{f_{\circ}}{u} \leqslant i$
and vertices $(\atR, v)$ with $\Slayer{f_{\circ}}{v} < i$.
The proof is by induction on the length of the path;
the base case is a path of length 0, which only visits a single prefix vertex,
and this vertex is on layer 0, so the statement is correct.

For the induction step, consider the second-to-last vertex on a path.
If it is $(\atL, u)$, then by inductive assumption $\Player{f_{\circ}}{u} \leqslant i$.
Let $(\atR, v)$ be the last vertex on the path.
Then, $v \in f(u)$,
and since $f$ drops down from the layer $i$, the last vertex has $\Slayer{f_{\circ}}{v} < i$.

Similarly, if the last two vertices on the path are $(\atR, v)$ and $(\atL, u)$,
then $\Slayer{f_{\circ}}{v} \leqslant i - 1$ by inductive assumption,
and $u \in g_I(v)$.
Then, by the definition of $g_I$,
the prefix layer of $u$ is $\Player{f_{\circ}}{u} \leqslant \Slayer{f_{\circ}}{v}+1 \leqslant i$,
as claimed.

Now, by Claim~\ref{A_of_g_I_claim},
the set $A(g_I)$ can only contain indices $q$ with $\Slayer{f_{\circ}}{q} \geqslant k - 1$.
On the other hand, every path in $G(f) \cup H(g_I)$
that starts in $(\atL, s(f))$ can only reach vertices $(\atR, v)$ 
with $\Slayer{f_{\circ}}{v} < i \leqslant k-1$.
Therefore, a path from $(\atL, s(f))$ cannot visit $(\atR, q)$ for any $q \in A(g_I)$,
and thus $K^{(n)}_{f,g_I}=0$.
\end{proof}

The last claim is in some sense complementary to Claim~\ref{sustenance_failing_claim}:
it says that if an ordered prefix table $f$ does not drop down from any layer of $f_{\circ}$,
then it affects the linear combination
on the columns corresponding to $g_I$, for various $I$,
and the elements $K^{(n)}_{f,g_I}$
depend only on the set $B_{f_{\circ}}(f)$ of breakthroughs of $f$,
and not on $f$ itself.

\begin{claim} \label{breakthrough_complement_claim}
Let $f$ and $f_{\circ}$ be two ordered prefix tables such that $f$ does not drop down from any layers of $f_{\circ}$. 
Let suffix tables $g_I$ be constructed using $f_{\circ}$ as described above.
Let $k = \rank f_{\circ}$.
Then, for every set $I \subseteq \{0, \ldots, k-1\}$,
$K^{(n)}_{f,g_I} = 1$ if and only if $I \cup B_{f_{\circ}}(f) = \{0, 1, \dots, k -1 \}$.
\end{claim}

\begin{proof}
Note that both $I$ and $B_{f_{\circ}}(f)$ are subsets of $\{0, 1, \dots, k -1 \}$.

\textcircled{$\Rightarrow$}
Suppose that there is an index $i \in \{0, 1, \dots, k -1 \}$
such that $i \notin I$ and $i \notin B_{f_{\circ}}(f)$, but $K^{(n)}_{f,g_I} = 1$.
Then, by Lemma~\ref{graph_acceptance_lemma},
there exists a path in $G(f) \cup H(g_I)$ that starts in $(\atL, s(f))$
and ends in $(\atR, q)$ for some $q \in A(g_I)$.

It will be shown that every path that begins in $(\atL, s(f))$
never goes beyond layer $i$ both on the left and on the right:
that is, it
only visits vertices $(\atL, u)$ with $\Player{f_{\circ}}{u} \leqslant i$
and vertices $(\atR, v)$ with $\Slayer{f_{\circ}}{v} \leqslant i$.
The proof uses induction on the path length;
the base case is a path of length 0,
which visits a single prefix vertex of layer 0.

For the induction step, consider the last two vertices on a path.
If the second-to-last vertex is $(\atL, u)$,
then by inductive assumption $\Player{f_{\circ}}{u} \leqslant i$.
The last vertex $(\atR, v)$ on the path satisfies $v \in f(u)$,
and since $f$ does not break through the $i$-th layer,
$\Slayer{f_{\circ}}{v} \leqslant i$.
Similarly, if the second-to-last vertex is $(\atR, v)$,
then by inductive assumption $\Slayer{f_{\circ}}{v} \leqslant i$,
and since $i \notin I$, then for all $u \in g_I(w)$
the layer $\Player{f_{\circ}}{u} \leqslant i$ by definition of $g_I$,
including the last vertex on the path.

By Claim~\ref{A_of_g_I_claim},
the set $A(g_I)$ can only contain indices $q$ with $\Slayer{f_{\circ}}{q} \geqslant k - 1$.
Since $i \leqslant k - 1$, a path from $(\atL, s(f))$ can visit $(\atR, q)$ with $q \in A(g_I)$ only if $i = k - 1$.
But then $k - 1 \notin I$,
and therefore $A(g_I)$ contains only indices $q$ with $\Slayer{f_{\circ}}{q} = k$,
so the path does not exist in this case as well.

\textcircled{$\Leftarrow$}
Now, suppose that the union of sets $I$ and $B_{f_{\circ}}(f)$ is $\{0, 1, \dots, k -1 \}$,
but $K^{(n)}_{f,g_I} = 0$.
Then, according to Lemma~\ref{graph_acceptance_lemma},
there in no path in the graph $G(f) \cup H(g_I)$ that starts in $(\atL, s(f))$
and ends in any vertex $(\atR, q)$ with $q \in A(g_I)$.

Let $U_0 = \{ u \mid (\atL, u) \text{ is reachable from } (\atL, s(f)) \text{ in } G(f) \cup H(g_I) \}$, 
and let $V_0 = \{ v \mid (\atR, v) \text{ is reachable from } (\atL, s(f)) \text{ in } G(f) \cup H(g_I) \}$.
Let $r_U = \max_{u \in U_0} \Player{f_{\circ}}{u}$,
and let $r_V = \max_{v \in V_0} \Slayer{f_{\circ}}{v}$.
Since no accepting vertices are reachable,
the set $V_0$ does not contain any indices from $A(g_I)$.
By Claim~\ref{A_of_g_I_claim}, this implies that $r_V \leqslant k - 1$,
and if $k - 1 \in I$, then furthermore $r_V \leqslant k - 2$.

It is claimed that $r_V=r_U$, which is proved as the following two inequalities.

\begin{description}
\item[($\bm{r_U \geqslant r_V}$)]
This inequality is inferred from the definition of $g_I$.
Indeed, let $v_0 \in V_0$ be the index with $\Slayer{f_{\circ}}{v_0} = r_V$.
Then, the vertex $(\atR, v_0)$ has arcs
leading into all vertices $(\atL, u)$ with $\Player{f_{\circ}}{u} \leqslant r_V$.
Hence, all those vertices are reachable.
Moreover, since $r_V \leqslant k - 1$ and the rank of $f_\circ$ is $k$,
there is at least one index $u$ with $\Player{f_{\circ}}{u} = r_V$,
as there is at least one $u$ with $f_{\circ}(u) = S_{r_V}^{f_{\circ}}$ by definition of $S^{f_{\circ}}$.
Accordingly, $r_U \geqslant \Player{f_{\circ}}{u} = r_V$.

\item[($\bm{r_V \geqslant r_U}$)]
To see that the inequality $r_V \geqslant r_U$ is also true,
first note that $U_0 = \{ u \mid \Player{f_\circ}{u} \leqslant r_U \}$.
This follows from the definition of $ g_I$ as follows.
If there is a path $P$ in the graph $G(f) \cup H(g_I)$ leading to some vertex $(\atL, u)$,
then there are paths to all vertices $(\atL, u')$
with $\Player{f_\circ}{u'} \leqslant \Player{f_\circ}{u}$,
obtained by replacing the last vertex on the path $P$;
those paths are valid by the definition of $g_I$.

Now, the table $f$ does not drop down from the layer $r_U$ of $f_{\circ}$,
that is, there are indices $u$ and $v$ such that $\Player{f_{\circ}}{u} \leqslant r_U$,
$\Slayer{f_{\circ}}{v} \geqslant r_U$, and $v \in f(u)$.
This $u$ is in $\{ u \mid \Player{f_\circ}{u} \leqslant r_U \}$,
and the latter set, as noted earlier, equals $U_0$.
Then, the vertex $(\atL, u)$ is reachable, and it has an arc leading to $(\atR, v)$.
Therefore, $(\atR, v)$ is also reachable, and $v \in V_0$.
Then, $r_V \geqslant \Slayer{f_{\circ}}{v} \geqslant r_U$.
\end{description}

As $r_V$ is both no less and no greater than $r_U$, those two values are equal: $r_V = r_U$.
Since $I \cup B_{f_{\circ}}(f) = \{0, 1, \dots, k -1 \}$ by the assumption,
and $r_U = r_V \in \{0, 1, \dots, k -1\}$,
this implies that either $r_U = r_V \in I$ or $r_U \in B_{f_{\circ}}(f)$.
Consider these two cases.

If $r_V \in I$, then the above argument for $r_U \geqslant r_V$
can be strengthened to show that $r_U \geqslant r_V + 1$.
Indeed, as mentioned above, $r_V \in I$ implies that $r_V \leqslant k - 2$,
that is, $r_V + 1 < k$,
and then, by the definition of $g_I$,
there are arcs from $(\atR, v_0)$
to all vertices $(\atL, u)$ with $\Player{f_{\circ}}{u} \leqslant r_V + 1$.
And since $\rank f_\circ = k > r_V + 1$, there is an index $u$ with $\Player{f_{\circ}}{u} = r_V + 1$.

If $r_U \in B_{f_{\circ}}(f)$, then the argument for $r_V \geqslant r_U$
is extended to prove that $r_V \geqslant r_U + 1$.
The condition that $f$ breaks through the layer $r_U$ of $f_{\circ}$
is that there exist indices $u$ and $v$ satisfying $\Player{f_{\circ}}{u} \leqslant r_U$,
$\Slayer{f_{\circ}}{v} > r_U$, and $v \in f(u)$.
As in the previous proof, $(\atR, v)$ is reachable via $(\atL, v)$,
and $v \in V_0$.
This time, $r_V \geqslant \Slayer{f_{\circ}}{v} \geqslant r_U + 1$.

Altogether, either $r_U \geqslant r_V + 1$ or $r_V \geqslant r_U + 1$,
and as $r_U$ and $r_V$ were proved to be equal,
both cases result in a contradiction.
This proves that $K^{(n)}_{f,g_I} = 1$, as claimed.
\end{proof}

Due to linear dependence, for every $I \subseteq \{0, 1, \dots, k -1\}$,
\begin{equation*}
	\sum_{f \text{ is an ordered prefix table}} \alpha_f K^{(n)}_{f,g_I} = 0
\end{equation*}
For every $f$ that drops down from some layer of $f_{\circ}$,
by Claim~\ref{sustenance_failing_claim},
the value of $K^{(n)}_{f,g_I}$ is $0$ for all $I$,
and so the terms for all such $f$ can be excluded from the sum.
\begin{equation*}
	\sum_{\substack{f \text{ is an ordered prefix table} \\
	\text{that does not drop down from any layers}
	}}
	\alpha_f K^{(n)}_{f,g_I}
		=
	0
\end{equation*}
All remaining $f$ do not drop down from any layers,
hence by Claim~\ref{breakthrough_complement_claim},
the value of $K^{(n)}_{f,g_I}$ is equal to $1$
if and only if $I \cup B_{f_{\circ}}(f) = \{0, 1, \dots, k -1 \}$,
and otherwise it is $0$.
Thus, it has been proved that
\begin{equation*}
	\sum_{\substack{f \text{ is an ordered prefix table} \\
	\text{that does not drop down from any layers}, \\
	\text{and } I \cup B_{f_{\circ}}(f) = \{0, 1, \dots, k -1 \}
	}}
	\alpha_f
		=
	0
\end{equation*}
Next, let $\alpha_f$ for ordered prefix tables $f$ with the same set of breakthroughs be grouped together.
For every set of breakthroughs $J \subseteq \{0, 1, \dots, k -1 \}$,
let $\alpha_J$ be the sum of $\alpha_f$ for all $f$ with $B_{f_{\circ}}(f) = J$
which do not drop down from any layers of $f_{\circ}$.
Then the above sum is expressed as
\begin{align*}
	\sum_{J \colon I \cup J = \{0, 1, \dots, k -1 \}} \alpha_J &= 0,\
		\quad \text{where} \\
	\alpha_J &= \sum_{\substack{f \text{ is an ordered prefix table} \\
	\text{that does not drop down from any layers}, \\
	\text{and } B_{f_{\circ}}(f)=J
	}}
	\alpha_f
\end{align*}
The left-hand side is equivalently expressed as 
$\sum_{J \supseteq \{0, 1, \dots, k -1 \} \setminus I} \alpha_J$,
where $\{0, 1, \dots, k -1\} \setminus I$ can be any subset of $\{0, 1, \dots, k -1\}$.
Therefore, for every set $\overline{I} \subseteq \{0, 1, \dots, k -1 \}$,
the following equality holds.
\begin{equation*}
	\sum_{J \supseteq \overline{I}} \alpha_J = 0
\end{equation*}

The following representing of $\alpha_J$ by the inclusion-exclusion principle
contains the above sum as a subexpression,
and hence the latter equality implies that every $\alpha_J$ is zero.

\begin{claim}
For every set $J \subseteq \{0, 1, \dots, k -1 \}$,
$\alpha_J = \sum_{J' \supseteq J} (-1)^{|J'| - |J|} \sum_{J'' \supseteq J'} \alpha_{J''}$.
In particular, $\alpha_J = 0$ for all $J$.
\end{claim}

\begin{proof}
For $J'' \supseteq J$,
let $\beta_{J''}$ be the total coefficient of $\alpha_{J''}$
in the above sum, so that it is represented as
$\sum_{J' \supseteq J} (-1)^{|J'| - |J|} \sum_{J'' \supseteq J'} \alpha_{J''}
	=
\sum_{J'' \supseteq J} \beta_{J''} \alpha_{J''}$.

If $J'' = J$, then $\beta_{J''} = 1$, since $\alpha_J$
occurs only once with coefficient 1
(that is, for $J'=J$ in the first sum, and $J''=J'$ in the second sum).

Now let $J'' \neq J$.
Then there exists an index $i$ such that $i \in J''$, but $i \notin J$.
Let us split all possible $J'$ with $J \subseteq J' \subseteq J''$ into pairs
which differ by the inclusion of $i$
(so, if $i \notin J'$, it is paired with $J' \cup \{ i \}$).
Each pair produces $\alpha_{J''}$ once with the coefficient $1$,
and once with the coefficient $-1$, which cancel out.
Therefore, in this case, $\beta_{J''} = 0$.

Hence, $\sum_{J' \supseteq J} (-1)^{|J'| - |J|} \sum_{J'' \supseteq J'} \alpha_{J''} = \alpha_J$,
as claimed.
Furthermore, by the above proof, every inner sum $\sum_{J'' \supseteq J'} \alpha_{J''}$ is zero,
and therefore, the whole sum is equal to $0$.
\end{proof}

On the other hand, $\alpha_J$ for $J=\emptyset$ must be non-zero
by another line of reasoning.

\begin{claim}
$\alpha_{\emptyset} = \alpha_{f_{\circ}}$. In particular, $\alpha_{\emptyset} \neq 0$.
\end{claim}

\begin{proof}
By definition, $\alpha_{\emptyset}$ is a sum of $\alpha_f$ such that $B_{f_{\circ}}(f) = \emptyset$,
and $f$ does not drop down from any layers of $f_{\circ}$.
Note that $f_{\circ}$ satisfies both conditions due to how layers are defined,
and is therefore included in the sum.

Suppose now that there is some other $f \neq f_{\circ}$ with $\alpha_f \neq 0$ which is included in the sum.
Due to the choice of $f_{\circ}$, the size $|f|$ is at least $|f_{\circ}|$.
Then, by Lemma~\ref{divergence_breakthrough_lemma},
the set $B_{f_{\circ}}(f)$ is not empty, which leads to a contradiction.

This confirms that $\alpha_{\emptyset} = \alpha_{f_{\circ}}$.
Since the coefficient $\alpha_{f_{\circ}}$ is non-zero by the choice of $f_{\circ}$,
it follows that $\alpha_{\emptyset} \neq 0$.
\end{proof}

Now the value of $\alpha_{\emptyset}$ cannot be simultaneously zero and non-zero.
Therefore, there is no non-trivial linear combination
of the form $\sum_{f} \alpha_f K^{(n)}_f = \mathbf{0}$,
and the rows of $K^{(n)}$ are linearly independent.
This completes the proof of Theorem~\ref{linear_independence_of_K_theorem}.
\end{proof}

This theorem gives a way to calculate the rank of the original matrix.

\begin{corollary} \label{unordered_matrix_rank_corollary}
$\rank M^{(n)}$ is equal to the number of ordered prefix tables on $[n]$.
\end{corollary}

\begin{proof}
By Theorem~\ref{matrix_reduction_theorem}, $\rank M^{(n)} = \rank K^{(n)}$.
By Theorem~\ref{linear_independence_of_K_theorem}, the rows of $K^{(n)}$ are linearly independent,
therefore, the rank of $K^{(n)}$ is equal to the number of rows in $K^{(n)}$.
By definition, the rows of $K^{(n)}$ are indexed by ordered prefix tables on $[n]$.
\end{proof}

\subsection{Rank calculation}

Corollary~\ref{unordered_matrix_rank_corollary} allows the rank of the original matrix
to be calculated using combinatorial methods.
The question is, how many ordered prefix tables on $[n]$ are there for a given $n$?

Let $f$ be an ordered prefix table.
Then, by definition of a prefix layer, $f(u) = S^f_{\Player{f}{u}}$ for every $u \in [n]$.
Hence, the \emph{prefix layer function} $\PL{f}$ mapping each $u \in [n]$ to its prefix layer,
and the sets $S_i^f \subseteq [n]$ of prefix layer contents, for all $0 \leqslant i \leqslant \rank f$,
together uniquely determine $f$.

However, there are some limitations on both objects.
The condition on the prefix layer function
is that every layer (except for the optional last one) actually exists,
that is, occurs in the image of $\PL{f}$.
This condition is expressed in the following abstract definition of a prefix layer function
that is independent of $f$.

\begin{definition}
Let $k$ be a non-negative integer.
A function $p \colon [n] \rightarrow \{0, 1, \dots, k\}$
is called a \emph{valid prefix layer function} of rank $k$,
if for every value $i \in \{0, 1, \dots, k-1\}$ there is $u \in [n]$ such that $p(u) = i$.
Let $P(n,k)$ be the number of valid prefix layer functions of rank $k$.
\end{definition}

The conditions on the sets $S_i^f$ is that they are non-empty and strictly nested within each other.

\begin{definition}
Let $k$ be a non-negative integer.
A sequence of sets $S_0 \subsetneq S_1 \subsetneq \dots \subsetneq S_k = [n]$
is called a \emph{valid nested set sequence} of rank $k$, if $S_0$ is non-empty.
Let $S(n, k)$ be the number of valid nested set sequences of rank $k$.
\end{definition}

It will now be proved that pairs of a valid prefix layer function and a valid nested set sequence
are in one-to-one correspondence with ordered prefix tables.
First, one can naturally extract such a pair from every ordered prefix table.

\begin{lemma} \label{prefix_table_to_valid_pair_lemma}
Let $f$ be an ordered prefix table, with $\rank f = k$.
Then, $\PL{f}$ is a valid prefix layer function of rank $k$,
and sequence $S_0^f, S_1^f, \dots, S_k^f$ is a valid nested set sequence of rank $k$.
\end{lemma}

\begin{proof}
For every $u \in [n]$, the value $\Player{f}{u}$ is the number of an actual layer. Therefore, $\Player{f}{u} \in \{0, 1, \dots, k\}$, and $\PL{f}$ maps $[n]$ to $\{0, 1, \dots, k \}$.

For every $i \in \{0, 1, \dots, k-1\}$, there is an index $u$ such that $f(u) = S_i^f$ (by definition of $S_i^f$). Then, $\Player{f}{u} = i$. Hence, $\PL{f}$ is a valid prefix layer function of rank $k$.

By definition, $S_0^f \subsetneq S_1^f \subsetneq \dots \subsetneq S_k^f = [n]$. Each of those is either equal to some $f(u)$, which is non-empty by the definition; or is equal to $[n]$, which is also non-empty. In particular, $S_0^f$ is non-empty. Therefore, $S_0^f, S_1^f, \dots, S_k^f$ is a valid nested set sequence of rank $k$.
\end{proof}

Next, it is proved that
this representation of an ordered prefix table into a valid prefix layer function
and a valid nested set sequence uniquely determines it.

\begin{lemma} \label{valid_pair_uniqueness_lemma}
Let $f$ and $f'$ be two ordered prefix tables such that $\PL{f} = \PL{f'}$ and $S_i^f = S_i^{f'}$ for all $i$. Then, $f = f'$.
\end{lemma}

\begin{proof}
By definition of a prefix layer, $f(u) = S^f_{\Player{f}{u}}$ and $f'(u) = S^{f'}_{\Player{f'}{u}}$ for all $u \in [n]$. Then, $f(u) = S^f_{\Player{f}{u}} = S^f_{\Player{f'}{u}} = S^{f'}_{\Player{f'}{u}} = f'(u)$ for all $u \in [n]$. Hence, $f = f'$.
\end{proof} 

Finally, every possible pair of those objects has a matching ordered prefix table.

\begin{lemma} \label{valid_pair_to_prefix_table_lemma}
Let $p$ be a valid prefix layer function of rank $k$, and let $S_0, S_1, \dots, S_k$ be a valid nested set sequence of rank $k$.
Define the function $f \colon [n] \rightarrow \mathcal{P}([n]) \setminus \emptyset$
by $f(u) = S_{p(u)}$.
Then, $f$ is an ordered prefix table, $\rank f = k$, $\PL{f} = p$, and $S_i^f = S_i$ for all $i \in \{0, 1, \dots, k\}$.
\end{lemma}

\begin{proof}
Since $\{S_i\}$ is a valid nested set sequence, the set $S_0$ is non-empty.
Since $S_0 \subseteq S_i$ for all $i$, the same is true for all sets $S_i$.
Since $S_i \subseteq S_k = [n]$ for all $i$, all sets $S_i$ are
nonempty subsets of $[n]$,
so $f$ is indeed a function from $[n]$ to $\mathcal{P}([n]) \setminus \emptyset$.

Since $p$ is a valid prefix function, there is an index $i \in [n]$ such that $p(i) = 0$
(if $k = 0$, it is the only possible value;
if $k > 0$, this follows from the definition).
Then, $f(i) = S_0$, and for every $j \in [n]$, $f(i)$ is a subset of $f(j)$.
Therefore, $i$ is a starting state, and $f$ satisfies the conditions on a prefix table.

Let $i, j \in [n]$ be two indices, and suppose that $p(i) \leqslant p(j)$.
Then, $f(i) = S_{p(i)} \subseteq S_{p(j)} = f(j)$.
Similarly, if $p(i) > p(j)$, then $f(j) \subseteq f(i)$.
Hence, for any two indices $i, j \in [n]$, either $f(i) \subseteq f(j)$ or $f(j) \subseteq f(i)$.
Therefore, by Lemma~\ref{ordered_prefix_table_lemma}, $f$ is an ordered prefix table.

Since $p$ is a valid prefix function, for every $i \in \{0, 1, \dots, k-1\}$ there is an index $u_i$ such that $p(u_i) = i$. Then, $f(u_i) = S_i$, and the set of all values of $f$ includes all $S_i$ (except, maybe, $S_k = [n]$). By definition of $f$, the set of all values of $f$ is included in the set of all $S_i$. Therefore, the sequence $S_i^f$ is obtained by ordering the set $\{S_0, S_1, \dots, S_{k-1}, S_k = [n]\}$ (since $[n]$ is always included in the resulting sequence), and all elements of this set are distinct. Hence, $S_i^f = S_i$. In particular, $\rank f = k$. 

Additionally, for any $u \in [n]$, the value $f(u) = S_{p(u)} = S_{p(u)}^f$, therefore $\Player{f}{u} = p(u)$. Since this is true for any $u$, the functions coincide: $\PL{f} = p$.
\end{proof}

This bijection provides the following way to calculate the number of ordered prefix tables of certain rank.

\begin{lemma} \label{valid_pair_and_prefix_table_bijection_lemma}
The number of ordered prefix tables on $[n]$ of rank $k$ is equal to $P(n, k) S(n, k)$.
\end{lemma}

\begin{proof}
By Lemmata~\ref{prefix_table_to_valid_pair_lemma},~\ref{valid_pair_uniqueness_lemma}
and~\ref{valid_pair_to_prefix_table_lemma},
there is a bijection between ordered prefix tables of rank $k$
and pairs consisting of a valid prefix layer function of rank $k$
and a valid nested set sequence of rank $k$.
Therefore, the sizes of sets of those objects are equal.
In addition, the number of pairs is equal
to the product of numbers of possible choices for each object in a pair.
Hence, the lemma is proved.
\end{proof}

Both limitations on valid objects can be expressed in terms of partitions.
The prefix layer function of rank $k$ can be seen as an ordered partition of $[n]$
into $k$ non-empty parts and one possibly empty part,
all of them pairwise disjoint, one for each layer number.
The nested set sequence of rank $k$ can be described
as an ordered partition of $[n]$ into $k + 1$ pairwise disjoint non-empty sets:
$S_0$ together with $k$ sets of differences $S_i \setminus S_{i-1}$.
This provides a way to express $P(n, k)$ and $S(n, k)$ in terms of partitions.

\begin{definition}
Let $k \leqslant n$ be integers.
A \emph{Stirling number of the second kind}, denoted $\stirling{n}{k}$,
is defined as the number of ways to partition the set $[n]$ into $k$ non-empty pairwise disjoint parts.
\end{definition}

It is well-known that the number of surjections from an $n$-element set to a $k$-element set
is exactly $k! \stirling{n}{k}$.

\begin{lemma} \label{valid_prefix_counting_lemma}
The number of valid prefix layer functions of rank $k$ is
$P(n, k) = k! \stirling{n+1}{k+1}$.
\end{lemma}
\begin{proof}
The proof is by establishing a bijection between
valid prefix layer functions of rank $k$
and surjective functions $\phi$ from $[n+1]$ to $[k+1]$ such that $\phi(n+1) = k+1$,
and then determining the number of such functions $\phi$.

Let $p \colon [n] \rightarrow \{0, 1, \dots, k\}$ be a valid prefix layer function of rank $k$,
that is, for every value $i \in \{0, 1, \dots, k-1\}$ there is an argument $u \in [n]$ with $p(u) = i$.
Consider a function $\phi_p \colon [n+1] \rightarrow [k+1]$
defined as $\phi_p(u) = p(u) + 1$ for $u \in [1, n]$, and $\phi_p(n+1) = k+1$.
Since all values between $0$ and $k-1$ belong to the image of $p$,
this function is surjective,
and it maps $n+1$ to $k+1$.
Furthermore, different valid prefix layer functions $p$ produce different $\phi_p$.
Therefore, there are at least as many different surjective functions with $\phi(n+1)=k+1$
as there are valid prefix layer functions of rank $k$.

To see that every surjective function $\phi \colon [n+1] \rightarrow [k+1]$ such that $\phi(n+1) = k+1$
can be expressed in the form of $\phi_p$ for some valid prefix layer function $p$ of rank $k$,
define the function $p \colon [n] \rightarrow \{0, \dots, k \}$ as $p(u) = \phi(u) - 1$.
It is enough to show that $p$ is a valid prefix layer function of rank $k$;
then, $\phi = \phi_p$ will hold true.
Since $\phi$ is surjective, for every $i \in \{0, \dots, k-1\}$,
there is an argument $u \in [n+1]$ such that $\phi(u) = i+1$.
Here $u$ cannot be $n+1$, because $i < k$, whereas $\phi(n+1) = k+1 \neq i+1$.
Then, $u \in [n]$, and accordingly, $p(u) = \phi(u) - 1 = i+1-1 = i$.
This establishes the desired bijection.

It remains to determine the number of surjective functions from $[n+1]$ to $[k+1]$ with $\phi(n+1) = k+1$.
For each $i \in [k+1]$, let $P(i)$ the number of surjective functions from $[n+1]$ to $[k+1]$
with $\phi(n+1) = i$.
By the symmetry, $P(1)=\ldots=P(n+1)$,
and since the total number of surjective functions is $\sum_i P(i)=(k+1)! \stirling{n+1}{k+1}$,
the desired number $P(n+1)$ is
$\frac{1}{k+1} \cdot (k+1)! \stirling{n+1}{k+1} = k! \stirling{n+1}{k+1}$.
\end{proof}

\begin{lemma} \label{valid_suffix_counting_lemma}
The number of valid nested set sequences of rank $k$ is
$S(n, k) = (k + 1)! \stirling{n}{k+1}$.
\end{lemma}

\begin{proof}
Let $S_{n,k}$ be the set of all valid nested set sequences on $[n]$ of rank $k$, and let $S_{n,k}'$ be the set of ordered partitions of $[n]$ into $k+1$ parts. Note that $|S_{n,k}'| = (k+1)! \stirling{n}{k+1}$, since an ordered partition is specified by a partition and an order of its blocks. Therefore, it is enough to show that $|S_{n,k}| = |S_{n,k}'|$.

Define a function $\phi \colon S_{n,k} \rightarrow S_{n,k}'$ as follows: for the valid nested set sequence $S$, consisting of sets $S_0, S_1, \dots, S_k = [n]$, define an ordered partition $B$, consisting of sets $B_0, B_1, \dots, B_k$, as $B_0 = S_0$ and $B_ i = S_i \setminus S_{i-1}$ for $i \in [k]$. Note that since $S$ is a valid nested set sequence, both $S_0$ and all of differences $S_i \setminus S_{i-1}$ are non-empty. Additionally, since $S_i$ are nested into each other, the sets $B_i$ do not intersect; also, since $S_k = [n]$, the union of all $B_i$ is equal to $[n]$ as well. Therefore, $B$ is an ordered partition of $[n]$ into $k + 1$ parts. Define $\phi(S) = B$.

Now, define an inverse function $\psi \colon S_{n,k}' \rightarrow S_{n,k}$ as follows: for the ordered partition $B$ of $[n]$ into sets $B_0, B_1, \dots, B_k$, define $S_i = \bigcup_{j=0}^i B_j$. Since $B_i$ are all non-empty, the sets $S_i$ are strictly nested into one another; also, $S_0 = B_0$ is non-empty and $S_k = \bigcup_{j=0}^k B_j = [n]$; therefore, the sequence $S$, composed of sets $S_i$, is a valid nested set sequence. Define $\psi(B) = S$.

The equality $\psi \circ \phi = \id_{S_{n, k}}$ follows from the fact that $S_i = S_0 \cup \bigcup_{j=1}^i S_j \setminus S_{j-1}$; the equality $\phi \circ \psi = \id_{P_{n, k}'}$ follows from the fact that $B_i = \bigcup_{j=0}^{i} B_j \setminus \bigcup_{j=0}^{i-1} B_j$. Therefore, $\phi$ and $\psi$ are both bijections. Hence, $|S_{n,k}| = |S_{n,k}'|$, which means that $S(n, k) = |S_{n,k}| = |S_{n,k}'| = (k + 1)! \stirling{n}{k+1}$.
\end{proof}

Those results can be combined into the final one.

\begin{theorem} \label{ordered_prefix_table_count_theorem}
There are $\sum_{k=0}^{n - 1} k! (k+1)! \stirling{n}{k+1} \stirling{n+1}{k+1} = \sum_{k=1}^{n} (k - 1)! k! \stirling{n}{k} \stirling{n+1}{k}$ different ordered prefix tables on $[n]$.
\end{theorem}

\begin{proof}
An ordered prefix table on $[n]$ can have a rank ranging from $0$ to $n - 1$, inclusive (since, for an ordered prefix table $f$, the size of a final set $|S_{\rank f}^f| = n$, but $|S_i^f| \geqslant i + 1$ due to being strictly contained into one another). By Lemma~\ref{valid_pair_and_prefix_table_bijection_lemma}, for each rank $k$, there are $P(n, k) S(n, k)$ ordered prefix tables of this rank. Therefore, there are $\sum_{k=0}^{n - 1} P(n, k) S(n, k)$ ordered prefix tables on $[n]$ in total.

By Lemma~\ref{valid_prefix_counting_lemma}, the value $P(n, k)$ is equal to $k! \stirling{n+1}{k+1}$. By Lemma~\ref{valid_suffix_counting_lemma}, the value $S(n, k) = (k + 1)! \stirling{n}{k+1}$. Substituting them into the formula above yields the desired result.
\end{proof}

\begin{theorem}
For every $n$, there is a language recognized by a 2NFA with $n$ states,
such that every UFA for the same language requires
at least $\sum_{k=1}^{n} (k - 1)! k! \stirling{n}{k} \stirling{n+1}{k}$ states.
\end{theorem}
\begin{proof}
By Theorem~\ref{Schmidt_theorem} for a 2NFA $\mathcal{A}_n$ and its language $L_n$, every UFA recognizing the same language has at least $\rank M^{(n)}$ states. By Corollary~\ref{unordered_matrix_rank_corollary}, the rank of $M^{(n)}$ is equal to the number of ordered prefix tables on $[n]$. Finally, by Theorem~\ref{ordered_prefix_table_count_theorem}, there are $\sum_{k=1}^{n} (k - 1)! k! \stirling{n}{k} \stirling{n+1}{k}$ ordered prefix tables on $[n]$. Therefore, every UFA recognizing the language $L_n$ requires at least $\sum_{k=1}^{n} (k - 1)! k! \stirling{n}{k} \stirling{n+1}{k}$ states.
\end{proof}

\subsection{Asymptotics of the lower bound}

\begin{theorem}
$\sum_{k=1}^{n} (k - 1)! k! \stirling{n}{k} \stirling{n+1}{k} = \Omega \left( \frac{n^{2n+2}}{e^{2n}} \right)$.
\end{theorem}
\begin{proof}
An asymptotic bound on this sum is obtained by estimating a single term, with $k = n$. Then, $\sum_{k=1}^{n} (k - 1)! k! \stirling{n}{k} \stirling{n+1}{k} \geqslant (n - 1)! n! \stirling{n}{n} \stirling{n+1}{n}$.

The Stirling numbers of the second kind in this term can be calculated precisely. 
There is only one partition of $[n]$ into $n$ parts, since then every part consists of a single element (hence, $\stirling{n}{n} = 1$).
Also, there are ${n \choose 2} = \frac{n(n-1)}{2}$ partitions of $[n]$ into $n - 1$ parts, since all parts except one must consist of a single element, and there are ${n \choose 2}$ ways to choose the remaining two-element part.
Therefore, $(n - 1)! n! \stirling{n}{n} \stirling{n+1}{n} = \frac{n(n-1)}{2} (n-1)! n! = \frac{n-1}{2} (n!)^2$.

By Stirling's approximation, $n! \sim \sqrt{2\pi n} \cdot \frac{n^n}{e^n}$. Hence, $\frac{n-1}{2} (n!)^2 \sim \frac{n}{2} \cdot 2 \pi n \cdot \frac{n^{2n}}{e^{2n}} = \pi \cdot \frac{n^{2n+2}}{e^{2n}} = \Omega \left( \frac{n^{2n+2}}{e^{2n}} \right)$.
\end{proof}

How does it fare compared to the other bounds on 2NFA transformation?
The transformation of 2UFA into UFA, which provies a natural lower bound for a 2NFA $\to$ UFA transformation,
has a proven lower bound of
$\sum_{k=1}^n {n \choose k - 1} {n \choose k} {2k - 2 \choose k - 1} = O(9^n \cdot n^{-3/2})$
states~\cite{SPetrovFPetrovOkhotin}, which grows merely exponentially.
The 2NFA-to-DFA transformation (an upper bound), however,
requires $\sum_{i=1}^n \sum_{j=1}^n {n \choose i} {n \choose j} (2^{n-i}-1)^{n-j}$ states,
which is of the order of $2^{n^2-O(n)}$.
The comparison between those three bounds is presented in the Table~\ref{tab:bounds_for_small_N}.

\section{Optimality of the lower bound} \label{bound_optimality_section}

There is a companion result that it is not possible to achieve a better lower bound via Schmidt's theorem by choosing another 2NFA and different pairs of strings.

\begin{theorem} \label{rank_optimality_theorem}
Let $\mathcal{A}$ be a 2NFA over an alphabet $\Sigma$ with $n$ states that recognizes a regular language $L$. 
Let $X = \{x_1, x_2, \dots, x_\ell\}$ and $Y = \{y_1, y_2, \dots, y_m\}$ be sets of strings over the alphabet $\Sigma$.
Let $M$ be a $\ell \times m$ matrix defined by $M_{i,j} = 1$ if $x_i y_j \in L$, and $M_{i,j} = 0$ otherwise.
Then, $\rank M \leqslant \sum_{k=1}^n (k - 1)! k! \stirling{n}{k} \stirling{n+1}{k}$.
\end{theorem}

In order to prove it, we will first introduce terminology
similar to that used in Section~\ref{lower_bound_section},
and establish results showing that the same conditions must be satisfied.

Let $Q = \{q_1, q_2, \dots, q_n\}$ be the set of states of the 2NFA $\mathcal{A}$. Let $Q_0$ be the set of starting states of $\mathcal{A}$.

In order to analyze the computations of $\mathcal{A}$ on all concatenations $x_iy_j$, we will consider all possible computations on prefixes $x_i$ and on suffixes $y_j$. The set of possible computations can be represented with prefix and suffix tables as follows.

For each prefix $x$ of a possible input string, the computations of $\mathcal{A}$ on the prefix $\lmark x$ can be represented by a function similar to a prefix table.

\begin{definition}
Let $x \in \Sigma^*$ be a string. Define the function $f_x \colon [n] \to \mathcal{P}([n])$ representing the computations of $\mathcal{A}$ on the input $\lmark x$ as follows. 
Let $S_x$ be the set of indices $i$ such that there is a computation of $\mathcal{A}$ on the string $\lmark x$ that starts in a state from $Q_0$ at the first symbol of $\lmark x$, and ends up moving to the right beyond the last symbol of $\lmark x$ into the state $q_i$.
In addition, for every $q_i \in Q$, let $T_{x,q_i}$ be the set of indices $j$ such that there is a computation of $\mathcal{A}$ on the string $\lmark x$ that starts in a state $q_i$ on the last symbol of $\lmark x$, and ends up moving to the right beyond the last symbol of $\lmark x$ into the state $q_j$.
Then, for every $u \in [n]$, define $f_x(u) = S_x \cup T_{x,q_u}$.
\end{definition}

In essence, the function $f_x$ is a transition function on the string $\lmark x$, but the transitions to states in $S_x$ are added to every possible state. Since the states from $S_x$ are reachable from start anyway, this should not change any acceptance status.

The function $f_x$ becomes a prefix table if a simple condition on $x$ is satisfied.

\begin{lemma} \label{arbitrary_prefix_table_connection_lemma}
Let $x \in \Sigma^*$ be a string such that $S_x$ is not empty.
Then, $f_x$ is a prefix table, and the value of $f_x$ on each starting state of $f_x$ is $S_x$.
\end{lemma}

\begin{proof}
Since $S_x$ is not empty, and for every $u \in [n]$ the value $f_x(u)$ contains $S_x$, all the values of $f_x$ are non-empty.
Therefore, $f_x$ is a function from $[n]$ to $\mathcal{P}([n]) \setminus \emptyset$.

To prove that $f_x$ is a prefix table, it remains to show that there is a starting state of $f_x$: that is, that there is an index $i$ such that $f_x(i) \subseteq f_x(j)$ for every $j \in [n]$. 

For that, let $S$ be the set of indices $i$ such that there is a computation of $\mathcal{A}$ on the string $\lmark x$ that starts in a state from $Q_0$ at the first symbol of $\lmark x$, and first enters the last symbol of $\lmark x$ in the state $q_i$. Note that $S_x \supseteq \bigcup_{i \in S} T_{x,q_i}$, since any computation from the definition of $S$ that ends in a state $q_i$ can be extended by a computation from the definition of $T_{x,q_i}$ to obtain a computation from the definition of $S_x$. Similarly, $S_x \subseteq \bigcup_{i \in S} T_{x,q_i}$, as any computation from the definition of $S_x$ can be split into two computations: one from the definiton of $S$ until the first visit of the last symbol of $\lmark x$, and the remainder from the definition of some $T_{x,q_i}$ for $i \in S$. Hence, $S_x = \bigcup_{i \in S} T_{x,q_i}$.

Since $S_x$ is not empty, and $S_x = \bigcup_{i \in S} T_{x,q_i}$, the set $S$ is not empty as well.
Let $i$ be any index in $S$.
Then $f_x(i) = S_x \cup T_{x,q_i} = S_x$ by the definition of $f_x$.
Hence, this index $i$ is a starting state,
because $f_x(i) = S_x \subseteq f_x(j)$ for every $j \in [n]$.

Since the values of a prefix table for all starting states are equal,
$f_x(j) = f_x(i) = S_x$ for each starting state $j$ of $f_x$.
\end{proof}

If those conditions are not satisfied, then the corresponding row of the matrix $M$ contains only zeroes, as is shown by the following lemma.

\begin{lemma} \label{arbitrary_prefix_table_triviality_lemma}
Let $x \in \Sigma^*$ be a string such that $S_x$ is empty. Then, for every string $y \in \Sigma^*$, their concatenation $xy$ is not in $L$.
\end{lemma}
\begin{proof}
Suppose the converse is true, and there is a string $y$ such that $xy \in L$. Consider an accepting computation $P$ of $\mathcal{A}$ on string $\lmark x y \rmark$. By definition, the automaton $\mathcal{A}$ can accept the string only while reading the symbol $\rmark$. Therefore, this accepting computation leaves the string $\lmark x$ at some point. Let $q_i$ be the state in which the automaton following the computation $P$ ends up after moving beyond the string $\lmark x$ for the first time. 

Let $P'$ be the computation obtained from $P$ by removing all steps after leaving the string $\lmark x$ for the first time. 
Then, the computation $P'$ can be seen as a computation on the string $\lmark x$ that starts in a state from $Q_0$ at the first symbol of $\lmark x$, and ends up moving to the right beyond the last symbol of $\lmark x$ into the state $q_i$.
Hence, by definition of $S_x$, the index $i$ is in $S_x$, which leads to a contradiction with $S_x$ being empty.
\end{proof}

The same results can be formulated for suffixes as well. 
For each suffix $y$ of a possible input string, the computations of $\mathcal{A}$ on the suffix $y \rmark$ can be represented by a function similar to a suffix table. 

Let $F$ be the set of accepting states of $\mathcal{A}$.

\begin{definition}
Let $y \in \Sigma^*$ be a string. Define the function $g_y \colon [n] \to \mathcal{P}([n] \cup \{ \Accept \})$ representing the computations of $\mathcal{A}$ on the input $y \rmark$ as follows. 
Let $A_y$ be the set of indices $i$ such that there is a computation of $\mathcal{A}$ on the string $y \rmark$ that starts in a state $q_i$ at the first symbol of $y \rmark$, and ends up in a state from $F$ on the last symbol of $y \rmark$.
In addition, for every $q_j \in Q$, let $T'_{y,q_j}$ be the set of indices $i$ such that there exists a computation of $\mathcal{A}$ on the string $y \rmark$ that starts in a state $q_j$ at the first symbol of $y \rmark$, and ends up moving to the left beyond the first symbol of $y \rmark$ into the state $q_i$.
Then, for every $v \in [n]$, define $g_y(v) = T'_{y,q_v}$ if $v \notin A_y$, and $g_y(v) = [n] \cup \{ \Accept \}$ otherwise.
\end{definition}

In essence, the function $g_y$ is a transition function on the string $y \rmark$,
but with the addition of all possible transitions from states $q_i$ for $i \in A_y$.
Since the automaton in one of those states at the first symbol of $y \rmark$ eventually accepts anyway,
this modification should not change the acceptance status of any strings.

\begin{lemma} \label{arbitrary_suffix_table_conversion_lemma}
Let $y \in \Sigma^*$ be a string such that $A_y$ is not empty. Then, $g_y$ is a suffix table.
\end{lemma}

\begin{proof}
Note that $g_y$ is already a function from $[n]$ to $\mathcal{P}([n] \cup \{ \Accept \})$.
As per Definition~\ref{definition_suffix_table},
it has to be checked that the value of $g_y$ on some index contains $\Accept$,
and if it contains $\Accept$, it must contain everything else.

Let $v$ be some index in $A_y$; it exists, since $A_y$ is not empty. Then, $g_y(v) = [n] \cup \{ \Accept \}$ by definition of $g_y$, and therefore, $\Accept \in g_y(v)$. Hence, there is an index $i$ such that $\Accept \in g_y(i)$.

Let $i$ be any index with $\Accept \in g_y(i)$.
Then, $i \in A_y$, because otherwise $g_y(i) = T'_{y,q_i}$,
which is a subset of $Q$ and does not contain $\Accept$. Hence, $g_y(i) = [n] \cup \{ \Accept \}$ by definition of $g_y$.
\end{proof}

\begin{lemma} \label{arbitrary_suffix_table_connection_lemma}
Let $y \in \Sigma^*$ be a string such that $A_y$ is not empty. Then, $A_y = A(g_y)$.
\end{lemma}

\begin{proof}
By definition, $A(g_y)$ is the set of indices $i$ such that $\Accept \in g_y(i)$.
By definition of $g_y$, the set $A_y$ is the set of indices $i$ such that $\Accept \in g_y(i)$ -- that is, such that $i \in A(g_y)$.
Therefore, $A_y = A(g_y)$.
\end{proof}

Similarly, if this condition does not hold, the corresponding column is empty.

\begin{lemma} \label{arbitrary_suffix_table_triviality_lemma}
Let $y \in \Sigma^*$ be a string such that $A_y$ is empty. Then, for every string $x \in \Sigma^*$, their concatenation $xy$ is not in $L$.
\end{lemma}

\begin{proof}
Suppose the converse is true, and there is a string $x$ such that $xy \in L$. Consider an accepting computation $P$ of $\mathcal{A}$ on string $\lmark xy \rmark$. By definition, the automaton $\mathcal{A}$ can accept the string only while reading the symbol $\rmark$. Therefore, this accepting computation enters the string $y \rmark$ at some point. Let $q_i$ be the state in which the automaton following the computation $P$ ends up after moving into the string $y \rmark$ for the last time.

Let $P'$ be the computation obtained from $P$ by removing all steps before entering the string $y \rmark$ for the last time. 
Then, the computation $P'$ can be seen as a computation on the string $y \rmark$ that starts in a state $q_i$ at the first symbol of $y \rmark$ and ends up in a state from $F$ on the last symbol of $y \rmark$. Hence, by definition of $A_y$, the index $i$ should be in the set $A_y$, which leads to a contradiction with $A_y$ being empty.
\end{proof}

Therefore, a reduced matrix can be defined. Let $X'$ be a subset of $X$ consisting of all strings $x \in X$ such that $S_x$ is not empty. Let $Y'$ be a subset of $Y$ consisting of all strings $y \in Y$ such that $A_y$ is not empty. Let $M'$ be a submatrix of $M$, with rows indexed by strings $x \in X'$ and columns indexed by strings $y \in Y'$, with $M'_{x_i, y_j} = M_{i, j}$. This is equivalent to $M'_{x, y} = 1$ if $xy \in L$, and $M'_{x, y} = 0$ otherwise.

\begin{lemma}\label{first_matrix_reduction_lemma}
The matrix $M'$ can be obtained from $M$ by removing some of the all-zero rows and columns,
and hence $\rank M' = \rank M$.
\end{lemma}
\begin{proof}
The matrix $M'$ can be obtained from the matrix $M$ by deleting rows corresponding to $x_i \notin X'$ and columns corresponding to $y_j \notin Y'$. By Lemmata~\ref{arbitrary_prefix_table_triviality_lemma}~and~\ref{arbitrary_suffix_table_triviality_lemma} respectively, those rows and columns consist of all zeroes.

The removal of all-zero rows or columns does not reduce the rank of the matrix, therefore the ranks of $M$ and $M'$ are equal.
\end{proof}

Thanks to this lemma, it is enough to prove the bound on the rank of the matrix $M'$.

It turns out that, for strings $x$ and $y$,
knowing the pair of functions $f_x$ and $g_y$ is sufficient to determine
whether the concatenation $xy$ is accepted by the automaton $\mathcal{A}$.
To be precise, its acceptance is determined
by the existence of a path in the graph $G(f_x) \cup H(g_y)$.
The following lemma states that if $xy$ is accepted, then the graph contains a certain path.

\begin{lemma} \label{M_prime_implies_M_n_lemma}
Let $x, y \in \Sigma^*$ be two strings such that $S_x$ is not empty, $A_y$ is not empty, and $xy \in L$.
Then, the graph $G(f_x) \cup H(g_y)$ has a path
that starts in $(\atL, s(f_x))$ and ends in $(\atR, q)$ for some $q \in A(g_y)$.
\end{lemma}

\begin{proof}
Let $P$ be an accepting computation of $\mathcal{A}$ on input $\lmark xy \rmark$. Such computation exists, since $xy \in  L$.

Since the computation $P$ starts at the left end-marker, and finishes at the right end-marker due to its acceptance, the computation $P$ crosses the border between $\lmark x$ and $y \rmark$ at least once, and an odd number of times in total. Hence, the computation $P$ can be split into parts $P = R_1 P_1 R_2 \dots R_k P_k$ where $R_i$ is a computation contained inside the string $\lmark x$, and leaves it with the last move; $P_i$ is a computation contained inside the string $y \rmark$, and leaves it with the last move (except for $P_k$, which accepts); the last moves of each part are border crossings between $\lmark x$ and $y \rmark$. Let $p_i$ for $1 \leqslant i \leqslant k -1$ be the index of the last state in the computation $P_i$, into which it moves when it leaves the string $y \rmark$; let $r_i$ for $1 \leqslant i \leqslant k$ be defined the same for $R_i$.

Note that, by definition, $r_1 \in S_x$, since $R_1$ is a computation that starts in a valid starting state for $\mathcal{A}$ and leaves the string $\lmark x$ into the state $q_{r_1}$. Accordingly, $r_k \in A_y$, since $P_k$ is a computation that starts in a state $q_{r_k}$ at the first symbol of $y \rmark$ and accepts without leaving the string $y \rmark$.

Define $p_0 = s(f_x)$.
To finish the proof, it is enough to show that the graph $G(f_x) \cup H(g_y)$
contains the arcs from $(\atR, r_i)$ to $(\atL, p_i)$,
and from $(\atL, p_i)$ to $(\atR, r_{i+1})$, for all $i$.
Indeed, in this case the sequence $(\atL, p_0), (\atR, r_1), (\atL, p_1), \dots, (\atL, p_{k-1}), (\atR, r_k)$ is a path in $G(f_x) \cup H(g_y)$ that starts in $(\atL, s(f_x))$ and ends in $(\atR, q)$ for some $q \in A(g_y)$, since $r_k \in A_y$ and $A_y = A(g_y)$ by Lemma~\ref{arbitrary_suffix_table_connection_lemma}. 

For all $i > 0$, the existence of the arcs follows directly from the definitions of $f_x$ and $g_y$, since $r_{i+1} \in T_{x,p_i} \subseteq f_x(p_i)$ (the corresponding computation is $R_{i+1}$), and $p_i \in T'_{y,r_i} \subseteq g_y(r_i)$ (the corresponding computation is $P_i$). Finally, $r_1 \in S_x \subseteq f_x(p_0)$, and so the graph $G(f_x) \cup H(g_y)$ contains an arc from $(\atL, p_0)$ to $(\atR, r_1)$ as well.
\end{proof}

Conversely, if such a path is in the graph, then the concatenation $xy$ is accepted.

\begin{lemma} \label{M_n_implies_M_prime_lemma}
Let $x, y \in \Sigma^*$ be two strings such that $S_x$ is not empty, $A_y$ is not empty, and the graph $G(f_x) \cup H(g_y)$ has a path that starts in $(\atL, s(f_x))$ and ends in $(\atR, q)$ for some $q \in A(g_y)$. Then, $xy \in L$.
\end{lemma}

\begin{proof}
Since $G(f_x) \cup H(g_y)$ is a bipartite graph with parts $\{ \atL \} \times [n]$ and $\{ \atR \} \times [n]$, any path in it alternates between those two parts.
Let $P$ be the shortest path in $G(f_x) \cup H(g_y)$ from $(\atL, s(f_x))$ to $(\atR, q)$ with $q \in A(g_y)$; let the vertices of $P$ be $(\atL, p_0), (\atR, r_1), (\atL, p_1), \dots, (\atL, p_{k-1}), (\atR, r_k)$, where $p_0 = s(f_x)$ and $r_k = q$. Note that $r_i \in f_x(p_{i-1}) = S_x \cup  T_{x,p_{i-1}}$ and $p_i \in g_y(r_i)$ by the definition of graphs  $G(f_x)$ and $H(g_y)$ respectively. Also note that $r_k \in A(g_y)$; since $A_y = A(g_y)$ by Lemma~\ref{arbitrary_suffix_table_connection_lemma}, this means that $r_k \in A_y$.

By Lemma~\ref{arbitrary_prefix_table_connection_lemma}, the value of the function $f_x(s(f_x))$ is equal to $S_x$.
Since $r_1 \in f_x(p_0) = f_x(s(f_x))$ by definition of the graph $G(f_x)$, then $r_1 \in S_x$.
Additionally, since $P$ is the shortest such path,
for all $i > 1$ the index $r_i$ cannot be in $S_x$,
otherwise there is an arc from $(\atL, p_0)$ to $(\atR, r_i)$ in the graph $G(f_x) \cup H(g_y)$,
and the path can be shortened. Hence, for $i > 1$, $r_i \in T_{x,p_{i-1}}$ instead.
Accordingly, for all $i < k$, the index $r_i$ is not in $A_y$, otherwise a shorter path can be constructed by removing everything in $P$ after the vertex $(\atR, r_i)$. Therefore, for $1 \leqslant i \leqslant k-1$, the value $g(r_i) = T'_{s,r_i}$, and thus $p_i \in g(r_i) = T'_{s,r_i}$.

Let $R_1$ be a computation of $\mathcal{A}$ on the string $\lmark x$ that starts in a state from $Q_0$ on a symbol $\lmark$, and ends up moving out of the string $\lmark x$ into the state $q_{r_1}$; such computation exists, since $r_1 \in S_x$. Let $R_i$ for $2 \leqslant i \leqslant k$ be a computation of $\mathcal{A}$ on the string $\lmark x$ that starts in a state $q_{p_{i-1}}$ on the last symbol of $\lmark x$, and ends moving out of the string $\lmark x$ into the state $q_{r_i}$; such computation exists, since $r_i \in T_{x,p_{i-1}}$.

Let $P_i$ for $1 \leqslant i \leqslant k-1$ be a computation of $\mathcal{A}$ on the string $y \rmark$ that starts in a state $q_{r_i}$ at the first symbol of $y \rmark$, and ends up moving out of the string $y \rmark$ into the state $q_{p_i}$; such computation exists, since $p_i \in T'_{s,r_i}$. Let $P_k$ be a a computation of $\mathcal{A}$ on the string $y \rmark$ that starts in a state $q_{r_k}$ at the first symbol of $y \rmark$, and ends up accepting the string; since $r_k \in A_y$, such computation exists.

Then, the computation $R_1 P_1 R_2 \dots R_k P_k$ is an accepting computation of $\mathcal{A}$ on the string $\lmark xy \rmark$, and therefore, $xy \in L$.
\end{proof}

Lemmata~\ref{M_prime_implies_M_n_lemma}--\ref{M_n_implies_M_prime_lemma}
imply a connection between the matrices $M'$ and $M^{(n)}$
(defined in the beginning of Section~\ref{lower_bound_section}),
which is expressed in the following lemma.

\begin{lemma} \label{M_prime_equates_M_n_lemma}
Let $x \in X'$ and $y \in Y'$ be two strings. Then, $M'_{x,y} = M^{(n)}_{f_x,g_y}$.
\end{lemma}

\begin{proof}
Since $x \in X'$, the set $S_x$ is not empty. Since $y \in Y'$, the set $A_y$ is also not empty. Therefore, by Lemma~\ref{arbitrary_prefix_table_connection_lemma}, $f_x$ is a prefix table; also, by Lemma~\ref{arbitrary_suffix_table_conversion_lemma}, $g_y$ is a suffix table.

By definition, $M'_{x,y} = 1$ if $xy \in L$, and 0 otherwise. By definition, $M^{(n)}_{f_x,g_y} = 1$ if the graph $G(f_x) \cup H(g_y)$ has a path that starts in $(\atL, s(f_x))$ and ends in $(\atR, q)$ for some $q \in A(g_y)$, and $M^{(n)}_{f_x,g_y} = 0$ otherwise. By Lemmata~\ref{M_prime_implies_M_n_lemma}~and~\ref{M_n_implies_M_prime_lemma}, those two conditions are equivalent.
\end{proof}

However, the matrix $M'$ need not be a submatrix of $M^{(n)}$,
as there might be duplicate rows corresponding to strings $x, x' \in X'$ with $f_x = f_{x'}$, or duplucate columns corresponding to strings $y, y' \in Y'$ with $g_y = g_{y'}$.

Therefore, yet another reduced matrix is defined, which will be denoted $M''$. Let $X''$ be the set of all prefix tables $f_x$ for $x \in X'$, and let $Y''$ be the set of all suffix tables $g_y$ for $y \in Y'$. Then, define $M''_{f_x,g_y} = M'_{x,y}$ for all $x \in X'$, $y \in Y'$. This definition is correct, as by Lemma~\ref{M_prime_equates_M_n_lemma} the value of $M'_{x,y}$ depends solely on the functions $f_x$ and $g_y$.

\begin{lemma} \label{second_matrix_reduction_lemma}
The matrix $M''$ can be obtained from $M'$ by removing some of the duplicate rows and columns.
Accordingly, $\rank M'' = \rank M'$.
\end{lemma}

\begin{proof}
For a prefix table $f \in X''$, let $x_f \in X'$ be some string which has this prefix table: $f_{x_f} = f$.
Let $X^{\circ}$ be the set of all representatives $x_f$, for different $f \in X''$.
Similarly, for a suffix table $g \in Y''$, let $y_g \in Y'$ be some string which has this suffix table: $g_{y_g} = g$.  Let $Y^{\circ}$ be the set of all $y_g$.

Then, the matrix $M''$ can be obtained from $M'$
by removing rows corresponding to $x \notin X^{\circ}$ and columns corresponding to $y \notin Y^{\circ}$.
Each of those rows and columns removed is duplicate of one in the reduced matrix:
by Lemma~\ref{M_prime_equates_M_n_lemma},
the row corresponding to $x \notin X^{\circ}$ coincides with the row for $x_{f_x} \in X^{\circ}$,
and the column corresponding to $y \notin Y^{\circ}$ coincides with the column for $y_{g_y} \in Y^{\circ}$.

Since the removal of duplicate rows or columns does not reduce the rank of the matrix, the ranks of $M'$ and $M''$ are equal.
\end{proof}

\begin{lemma} \label{third_matrix_reduction_lemma}
The matrix $M''$ is a submatrix of $M^{(n)}$,
and therefore $\rank M'' \leqslant \rank M^{(n)}$.
\end{lemma}
\begin{proof}
By Lemma~\ref{M_prime_equates_M_n_lemma}, for all $x \in X'$ and $y \in Y'$, $M''_{f_x,g_y} = M'_{x,y} = M^{(n)}_{f_x,g_y}$. Hence, the matrix $M''$ is a submatrix of $M^{(n)}$.
\end{proof}

Those results together represent the proof of the theorem.

\begin{proof}[Proof of Theorem~\ref{rank_optimality_theorem}]
By Lemma~\ref{first_matrix_reduction_lemma}, $\rank M = \rank M'$.
By Lemma~\ref{second_matrix_reduction_lemma}, $\rank M' = \rank M''$.
By Lemma~\ref{third_matrix_reduction_lemma}, $\rank M'' \leqslant \rank M^{(n)}$.
By Corollary~\ref{unordered_matrix_rank_corollary}, the rank of $M^{(n)}$ is equal to the number of ordered prefix tables on $[n]$. 
Finally, by Theorem~\ref{ordered_prefix_table_count_theorem}, there are $\sum_{k=1}^{n} (k - 1)! k! \stirling{n}{k} \stirling{n+1}{k}$ ordered prefix tables on $[n]$.

Therefore, $\rank M \leqslant \sum_{k=1}^{n} (k - 1)! k! \stirling{n}{k} \stirling{n+1}{k}$.
\end{proof}

\section{Conclusion}

The lower bound on the state complexity of transforming 2NFA to UFA
established in this paper is of the order $\Omega \left( \frac{n^{2n+2}}{e^{2n}} \right)$,
which is well above the upper bound of $O(2^n \cdot n!)$
on the 2UFA-to-UFA transformation,
established in the previous paper by the authors~\cite{LATA}.
This result shows that the tradeoff functions for the 2NFA-to-UFA
and 2UFA-to-UFA transformations must be different.
However, the new lower bound is still far from the current upper bound,
derived from 2NFA-to-DFA transformation, which is of the order $2^{n^2 - O(n)}$~\cite{Kapoutsis_thesis}.
The lower bound derived in this paper
is compared to the known bounds on the 2NFA-to-UFA tradeoff
in Table~\ref{tab:bounds_for_small_N}.

All bounds presented in the table
rely on using an alphabet of exponential size in the number of states;
the bounds might be different in the case of alphabets of subexponential size,
as is known to be the case for 2DFA-to-DFA and 2DFA-to-NFA transformations~\cite{GeffertOkhotin,GeffertOkhotin_2dfa_to_1dfa}.

It would be interesting to determine the 2NFA-to-UFA tradeoff precisely.
However, as shown in Theorem~\ref{rank_optimality_theorem},
it is not possible to obtain a greater lower bound
by the means of Schmidt's theorem alone.
Hence, new methods would be needed for any further improvements to the lower bound. 

An improvement of the existing upper bound,
if possible at all, would require a completely new construction
showcasing the difference between UFAs and DFAs
as they are faced with simulating unrestricted nondeterminism.
No such construction exists in the case of one-way NFA,
as the NFA-to-UFA and the NFA-to-DFA tradeoffs
are practically the same ($2^n - 1$ versus $2^n$)~\cite{Leung}.
Hence, if any substantial difference could be found
between the 2NFA-to-UFA and 2NFA-to-DFA transformations,
this would be interesting indeed.

\begin{table}[h]
\caption{The lower bound established in this paper for small values of $n$,
	compared to the known tradeoffs from two-way to one-way finite automata.}
\label{tab:bounds_for_small_N}
\begin{center}
\begin{tabular}{|c|r|r|r|r|}
\hline
\multirow{4}{*}{$n$} & \multicolumn{1}{c|}{2DFA $\to$ UFA} & \multicolumn{1}{c|}{2DFA $\to$ UFA} & \multicolumn{1}{c|}{\textbf{2NFA $\to$ UFA}} & \multicolumn{1}{c|}{2NFA $\to$ DFA} \\
& \multicolumn{1}{c|}{2UFA $\to$ UFA} & \multicolumn{1}{c|}{2UFA $\to$ UFA} & \multicolumn{1}{c|}{(lower bound)} & \\
& \multicolumn{1}{c|}{(lower bound)} & \multicolumn{1}{c|}{(upper bound)} & & \\
\cline{2-5}
& \multicolumn{1}{c|}{$\sum\limits_{k=1}^{n} {n \choose k -1} {n \choose k} {2k - 2 \choose k - 1}$}
& \multicolumn{1}{c|}{$\sum\limits_{k=1}^{n} {n \choose k -1} {n \choose k} k!$}
& \multicolumn{1}{c|}{$\sum\limits_{k=1}^{n} (k - 1)! k! \stirling{n}{k} \stirling{n+1}{k}$}
& \multicolumn{1}{c|}{$\sum\limits_{i=1}^n \sum\limits_{j=1}^n {n \choose i} {n \choose j} (2^{n-i}-1)^{n-j}$} \\
\hline
1 & 1 & 1 & 1 & 1 \\
\hline
2 & 6 & 6 & 7 & 7 \\
\hline
3 & 39 & 39 & 115 & 133 \\
\hline
4 & 276 & 292 & 3\,451 & 7\,891 \\
\hline
5 & 2\,055 & 2\,505 & 164\,731 & 1\,613\,581 \\
\hline
6 & 15\,798 & 24\,306 & 11\,467\,387 & 1\,201\,168\,507 \\
\hline
7 & 124\,173 & 263\,431 & 1\,096\,832\,395 & 3\,360\,710\,751\,133 \\
\hline
8 & 992\,232 & 3\,154\,824 & 138\,027\,417\,451 & 36\,005\,748\,492\,454\,531 \\
\hline
\end{tabular}
\end{center}
\end{table}

\section*{Acknowledgements}

This work was supported by the Russian Science Foundation, project 23-11-00133.


\begin{thebibliography}{99}

\bibitem{Birget_state_compressibility} J.-C. Birget,
	\href{http://dx.doi.org/10.1007/BF01371727}
	{``State-complexity of finite-state devices, state compressibility and incompressibility''},
	\emph{Mathematical Systems Theory},
	26:3 (1993), 237--269.
	
\bibitem{Chrobak} M. Chrobak,
	\href{http://dx.doi.org/10.1016/0304-3975(86)90142-8}
	{``Finite automata and unary languages''},
	\emph{Theoretical Computer Science}, 
	47 (1986), 149--158.
	\href{http://dx.doi.org/10.1016/S0304-3975(03)00136-1}
	{Errata}:
	302 (2003), 497--498.
	
\bibitem{CzerwinskiDebskiGogaszHoiJainSkrzypczakStephanTan} W. Czerwi\'nski, M. D\k{e}bski, T. Gogasz, G. Hoi, S. Jain, M. Skrzypczak, F. Stephan, Ch. Tan,
	\href{https://doi.org/10.4230/LIPIcs.FSTTCS.2023.22}
	{``Languages given by finite automata over the unary alphabet''},
	\emph{43rd IARCS Annual Conference
	on Foundations of Software Technology and Theoretical Computer Science}
	(FSTTCS 2023, December 18--20, 2023, Hyderabad, India),
	LIPIcs 284,
	22:1--22:20.

\bibitem{GeffertMereghettiPighizzini2003} V. Geffert, C. Mereghetti, G. Pighizzini,
	\href{http://dx.doi.org/10.1016/S0304-3975(02)00403-6}
	{``Converting two-way nondeterministic unary automata into simpler automata''},
	\emph{Theoretical Computer Science},
	295:1--3 (2003), 189--203.

\bibitem{GeffertMereghettiPighizzini} V. Geffert, C. Mereghetti, G. Pighizzini,
	\href{http://dx.doi.org/10.1016/j.ic.2007.01.008}
	{``Complementing two-way finite automata''},
	\emph{Information and Computation},
	205:8 (2007), 1173--1187.

\bibitem{GeffertOkhotin} V. Geffert, A. Okhotin,
	{``One-way simulation of two-way finite automata over small alphabets''},
	\emph{NCMA 2013}
	(Ume{\aa}, Sweden, 13--14 August 2013).

\bibitem{GeffertOkhotin_2afa} V. Geffert, A. Okhotin,
	\href{http://dx.doi.org/10.1007/978-3-662-44522-8_25}
	{``Transforming two-way alternating finite automata to one-way nondeterministic automata''},
	\emph{Mathematical Foundations of Computer Science}
	(MFCS 2014, Budapest, Hungary, 25--29 August 2014),
	Part I, LNCS 8634, 291--302.

\bibitem{GeffertOkhotin_2dfa_to_1dfa} V. Geffert, A. Okhotin,
	\href{https://doi.org/10.1007/978-3-030-93489-7_3}
	{``Deterministic one-way simulation of two-way deterministic finite automata over small alphabets''},
	\emph{Descriptional Complexity of Formal Systems 2021},
	LNCS 13037, 26--37.
	
\bibitem{GoosKieferYuan} M. G\"o\"os, S. Kiefer, W. Yuan,
	\href{https://doi.org/10.4230/LIPIcs.ICALP.2022.126}
	{``Lower bounds for unambiguous automata via communication complexity''},
	\emph{49th International Colloquium on Automata, Languages, and Programming}
	(ICALP 2022, July 4--8, 2022, Paris, France),
	LIPIcs 229, 126:1--126:13.

\bibitem{IndzhevKiefer} E. Indzhev, S. Kiefer,
	\href{https://doi.org/10.1016/j.ipl.2022.106270}
	{``On complementing unambiguous automata and graphs with many cliques and cocliques''},
	\emph{Information Processing Letters},
	177 (2022), article 106270.

\bibitem{JirasekjrJiraskovaSebej} J. Jir\'asek Jr., G. Jir\'askov\'a, J. \v{S}ebej,
	\href{http://dx.doi.org/10.1142/S012905411842008X}
	{``Operations on unambiguous finite automata''},
	\emph{International Journal of Foundations of Computer Science},
	29:5 (2018), 861--876.

\bibitem{Kapoutsis} C. A. Kapoutsis,
	\href{http://dx.doi.org/10.1007/11549345_47}
	{``Removing bidirectionality from nondeterministic finite automata''},
	\emph{Mathematical Foundations of Computer Science}
	(MFCS 2005, Gdansk, Poland, 29 August--2 September 2005),
	LNCS 3618, 544--555.

\bibitem{Kapoutsis_thesis} C. A. Kapoutsis,
	\emph{Algorithms and Lower Bounds in Finite Automata Size Complexity},
	Ph.~D.\ thesis,
	Massachusetts Institute of Technology, 2006.

\bibitem{Kapoutsis_logspace} C. A. Kapoutsis,
	\href{https://doi.org/10.1007/s00224-013-9465-0}
	{``Two-way automata versus logarithmic space''},
	\emph{Theory of Computing Systems},
	55:2 (2014), 421--447.

\bibitem{KapoutsisPighizzini_logspace} C. A. Kapoutsis, G. Pighizzini,
	\href{http://dx.doi.org/10.1007/978-3-642-30642-6_21}
	{``Two-way automata characterizations of L/poly versus NL''},
	\emph{Theory of Computing Systems},
	56:4 (2015), 662--685.

\bibitem{TwoWayDFAs} M. Kunc, A. Okhotin,
	\href{http://dx.doi.org/10.1007/978-3-642-22321-1_28}
	{``Describing periodicity in two-way deterministic finite automata using transformation semigroups''},
	\emph{Developments in Language Theory}
	(DLT 2011, Milan, Italy, 19--22 July 2011),
	LNCS 6795, 324--336.
	
\bibitem{Leung1998} H. Leung,
	\href{https://doi.org/10.1137/S0097539793252092}
	{``Separating exponentially ambiguous finite automata from polynomially ambiguous finite automata''},
	\emph{SIAM Journal on Computing},
	27:4 (1998), 1073--1082.

\bibitem{Leung} H. Leung,
	\href{http://dx.doi.org/10.1142/S0129054105003418}
	{``Descriptional complexity of NFA of different ambiguity''},
	\emph{International Journal of Foundations of Computer Science},
	16:5 (2005), 975--984.

\bibitem{MereghettiPighizzini2001} C. Mereghetti, G. Pighizzini,
	\href{http://dx.doi.org/10.1137/S009753979935431X}
	{``Optimal simulations between unary automata''},
	\emph{SIAM Journal on Computing},
	30:6 (2001), 1976--1992.

\bibitem{ufa_sc} A. Okhotin,
	\href{http://dx.doi.org/10.1016/j.ic.2012.01.003}
	{``Unambiguous finite automata over a unary alphabet''},
	\emph{Information and Computation},
	212 (2012), 15--36.

\bibitem{FPetrov_ufa} F. Petrov,
	\href{https://doi.org/10.1007/s10474-023-01299-6}
	{``Logarithmic asymptotics of Landau--Okhotin function''},
	\emph{Acta Mathematica Hungarica},
	169 (2023), 272--276.

\bibitem{LATA} S. Petrov, A. Okhotin,
	\href{https://doi.org/10.1016/j.ic.2022.104956}
	{``On the transformation of two-way deterministic finite automata to unambiguous finite automata''},
	\emph{Information and Computation},
	295A (2023), article 104956.
	
\bibitem{SPetrovFPetrovOkhotin} S. Petrov, F. Petrov, A. Okhotin,
	\href{https://doi.org/10.48550/arXiv.2312.05909}
	{``On the rank of the communication matrix for deterministic two-way finite automata''},
	CoRR abs/2312.05909 (2023).

\bibitem{RadionovaOkhotin} M. Radionova, A. Okhotin,
	\href{http://dx.doi.org/10.4204/EPTCS.388.11}
	{``Sweeping permutation automata''},
	\emph{Non-Classical Models of Automata and Applications}
	(NCMA 2023, Famagusta, North Cyprus, 18--19 September 2023),
	EPTCS 388, 110--124.

\bibitem{Raskin} M. Raskin,
	\href{https://doi.org/10.4230/LIPIcs.ICALP.2018.138}
	{``A superpolynomial lower bound for the size of non-deterministic complement of an unambiguous automaton''},
	\emph{45th International Colloquium on Automata, Languages, and Programming}
	(ICALP 2018, Prague, Czech Republic, July 9--13, 2018),
	LIPIcs 107.

\bibitem{RazSpieker} R. Raz, B. Spieker,
	\href{https://doi.org/10.1007/BF01192528}
	{``On the `log rank'-conjecture in communication complexity''},
	\emph{Combinatorica},
	15:4 (1995), 567--588.

\bibitem{Schmidt} E. M. Schmidt,
	\emph{Succinctness of Description of Context-Free, Regular and Unambiguous Languages},
	Ph.\ D. thesis,
	Cornell University, 1978.

\bibitem{Shepherdson} J. C. Shepherdson,
	\href{http://dx.doi.org/10.1147/rd.32.0198}
	{``The reduction of two-way automata to one-way automata''},
	\emph{IBM Journal of Research and Development},
	3 (1959), 198--200.

\bibitem{Vardi} M. Vardi,
	\href{http://dx.doi.org/10.1016/0020-0190(89)90205-6}
	{``A note on the reduction of two-way automata to one-way automata''},
	\emph{Information Processing Letters},
	30:5 (1989), 261--264.

\end{thebibliography}
\end{document}